\documentclass[11pt]{article}
\usepackage[top=1in, bottom=1in, left=1in, right=1in]{geometry}
\usepackage{tgpagella}
\usepackage[utf8]{inputenc} 
\usepackage[T1]{fontenc}    

\usepackage{url}
\usepackage{hyperref}
\usepackage{amsmath,amsthm,amssymb,amsbsy,amsfonts,amscd,bm}
\usepackage{paralist}
\usepackage{xcolor}
\usepackage{color}
\usepackage{cleveref}
\usepackage{graphicx}
\graphicspath{{./figs/}}
\usepackage{algorithm}
\usepackage{algorithmic}
\usepackage{comment}
\usepackage{multirow}
\usepackage{enumitem}
\usepackage{fancyhdr}
\usepackage{subcaption}
\usepackage{wrapfig}

\theoremstyle{remark}

\newcommand{\e}{\begin{equation}}
\newcommand{\ee}{\end{equation}}
\newcommand{\en}{\begin{equation*}}
\newcommand{\een}{\end{equation*}}
\newcommand{\eqn}{\begin{eqnarray}}
\newcommand{\eeqn}{\end{eqnarray}}
\newcommand{\bmat}{\begin{bmatrix}}
\newcommand{\emat}{\end{bmatrix}}

\DeclareMathAlphabet\mathbfcal{OMS}{cmsy}{b}{n}

\DeclareMathOperator*{\argmin}{\text{arg~min}}

\newcommand{\calA}{\mathcal{A}}

\newcommand{\calM}{\mathcal{M}}

\graphicspath{{./figs/}}

\newlength{\imgwidth}
\newboolean{twoColVersion}
\setboolean{twoColVersion}{false}
\newcommand{\twoCol}[2]{\ifthenelse{\boolean{twoColVersion}} {#1} {#2} }

\newcommand*\samethanks[1][\value{footnote}]{\footnotemark[#1]}

\newcommand{\xmath}[1] {\ensuremath{#1}\xspace}

\newcommand{\blmath}[1] {\xmath{\bm{#1}}}

\newcommand{\x} {\blmath{x}}
\newcommand{\y} {\blmath{y}}
\newcommand{\blz} {\blmath{z}}
\newcommand{\blv} {\blmath{v}}
\newcommand{\blb} {\blmath{b}}
\newcommand{\cc} {\blmath{c}}
\newcommand{\X} {\blmath{X}}
\newcommand{\I} {\blmath{I}}
\newcommand{\W} {\blmath{W}}
\newcommand{\blR} {\blmath{R}}

\newcommand{\Ws} {\xmath{\W_s}}
\newcommand{\Rs} {\xmath{\blR_s}}
\newcommand{\bs} {\xmath{\blb_s}}
\newcommand{\cs} {\xmath{\cc_s}}
\newcommand{\Xs} {\xmath{\X_s}}
\newcommand{\Is} {\xmath{\tilde{\I}_s}}
\newcommand{\Izero} {\xmath{\tilde{\I}_0}}
\newcommand{\Ione} {\xmath{\tilde{\I}_1}}

\newcommand{\cA} {\xmath{\mathcal{A}}} 
\newcommand{\cB} {\xmath{\mathcal{B}}}
\newcommand{\cL} {\xmath{\mathcal{L}}} 
\newcommand{\cN} {\xmath{\mathcal{N}}} 
\newcommand{\cO} {\xmath{\mathcal{O}}} 

\newcommand{\xnext} {\xmath{\x_{k+1}}}
\newcommand{\xk} {\xmath{\x_k}}
\newcommand{\yk} {\xmath{\y_k}}
\newcommand{\ynext} {\xmath{\y_{k+1}}}

\newcommand{\der}{\, \mathrm{d}}
\newcommand{\As}{\xmath{\lambda_s}} 
\newcommand{\Frac}[2]{#1/#2}

\definecolor{customcolor}{RGB}{128, 0, 50}

\usepackage{authblk}
\usepackage{booktabs}
\usepackage{cleveref}
\usepackage{cite}
\PassOptionsToPackage{numbers, sort&compress}{natbib}
\usepackage{natbib}
\usepackage{mathtools}
\usepackage[toc, page]{appendix}
\usepackage{wrapfig}
\usepackage{hyperref}

\hypersetup{
    colorlinks=true,%
    citecolor=blue,%
    filecolor=blue,%
    linkcolor=blue,%
    urlcolor=blue
}

\usepackage{bm}
\usepackage{xspace}
\usepackage{tocloft}
\title{FlowDAS: A Stochastic Interpolant-based Framework for Data Assimilation}

\author[$\Diamond$,$\dagger$]{Siyi Chen\thanks{The first two authors contributed equally to the work. This work was completed while Siyi Chen was an intern at the University of Michigan.}}
\author[$\Diamond$]{Yixuan Jia\samethanks}
\author[$\Diamond$]{Qing Qu}
\author[$\mathsection$]{He Sun\thanks{Corresponding author: \texttt{hesun@pku.edu.cn}}}
\author[$\Diamond$]{Jeffrey A. Fessler}

\affil[$\Diamond$]{Department of Electrical Engineering and Computer Science, University of Michigan}
\affil[$\dagger$]{Department of Physics, Peking University}
\affil[$\mathsection$]{National Biomedical Imaging Center, Peking University}

\date{}

\begin{document}

\maketitle
\begin{abstract}

\noindent Data assimilation (DA) integrates observations with a dynamical model to estimate states of PDE-governed systems. Model-driven methods (e.g., Kalman, particle) presuppose full knowledge of the true dynamics, which is not always satisfied in practice, while purely data-driven solvers learn a deterministic mapping between observations and states and therefore miss the intrinsic stochasticity of real processes. 
Recently, Score-based diffusion models learn a global diffusion prior and provide a good modelling of the stochastic dynamics, showing new potential for DA. However, their all-at-once generation rather than the step-by-step transition limits their performance when dealing with highly complex stochastic processes and lacks physical interpretability. To tackle these drawbacks, we introduce \textbf{FlowDAS}, a generative DA framework that uses stochastic interpolants to directly learn state transition dynamics and achieve step-by-step transition to better model the real dynamics. We also improve the framework by combining the observation, better suiting the DA settings. Directly learning the underlying dynamics from collected data removes restrictive dynamical assumptions, and conditioning on observations at each interpolation step yields stable, measurement-consistent forecasts. Experiments on Lorenz-63, Navier–Stokes super-resolution/sparse-observation scenarios, and large-scale weather forecasting—--where dynamics are partly or wholly unknown---show that FlowDAS surpasses model-driven methods, neural operators, and score-based baselines in accuracy and physical plausibility. The source code is available at \url{https://github.com/umjiayx/FlowDAS.git}.

\end{abstract}
\tableofcontents
\vspace{-0.2in}
\section{Introduction}
\label{sec:intro}
\vspace{-0.1in}

Recovering state variables in complex dynamical systems is a fundamental problem in science and engineering. Accurate state estimation from noisy, incomplete data is critical in weather forecasting \cite{bocquet2015data,reichle2008data}, oceanography \cite{cummings2005operational,cummings2013variational}, seismology \cite{werner2009earthquake,banerjee2023parameter}, 
and many other fields \cite{wang2000data}, where reliable predictions depend on understanding the underlying physics \cite{bocquet2010beyond,bannister2017review,reich2015probabilistic}. A representative example is fluid dynamics 
\cite{taira2020model,brunton2015closed,duraisamy2019turbulence}, where one aims to reconstruct a continuous velocity field from sparse, noisy observations governed by nonlinear, time-dependent partial differential equations (PDEs)—a task complicated by stochasticity and high dimensionality. To meet these challenges, \textbf{data assimilation} (DA) combines model forecasts with observations to produce physically consistent state estimates; developed first in atmospheric and oceanic forecasting,
DA is now ubiquitous across many scientific and engineering domains \cite{rabier2005overview,geer2018all,gustafsson2018survey,rodell2004global,van2010nonlinear,fletcher2017data,carrassi2018data}. 

\begin{figure*}[t]  
\begin{center}
    \includegraphics[width=0.9\textwidth]{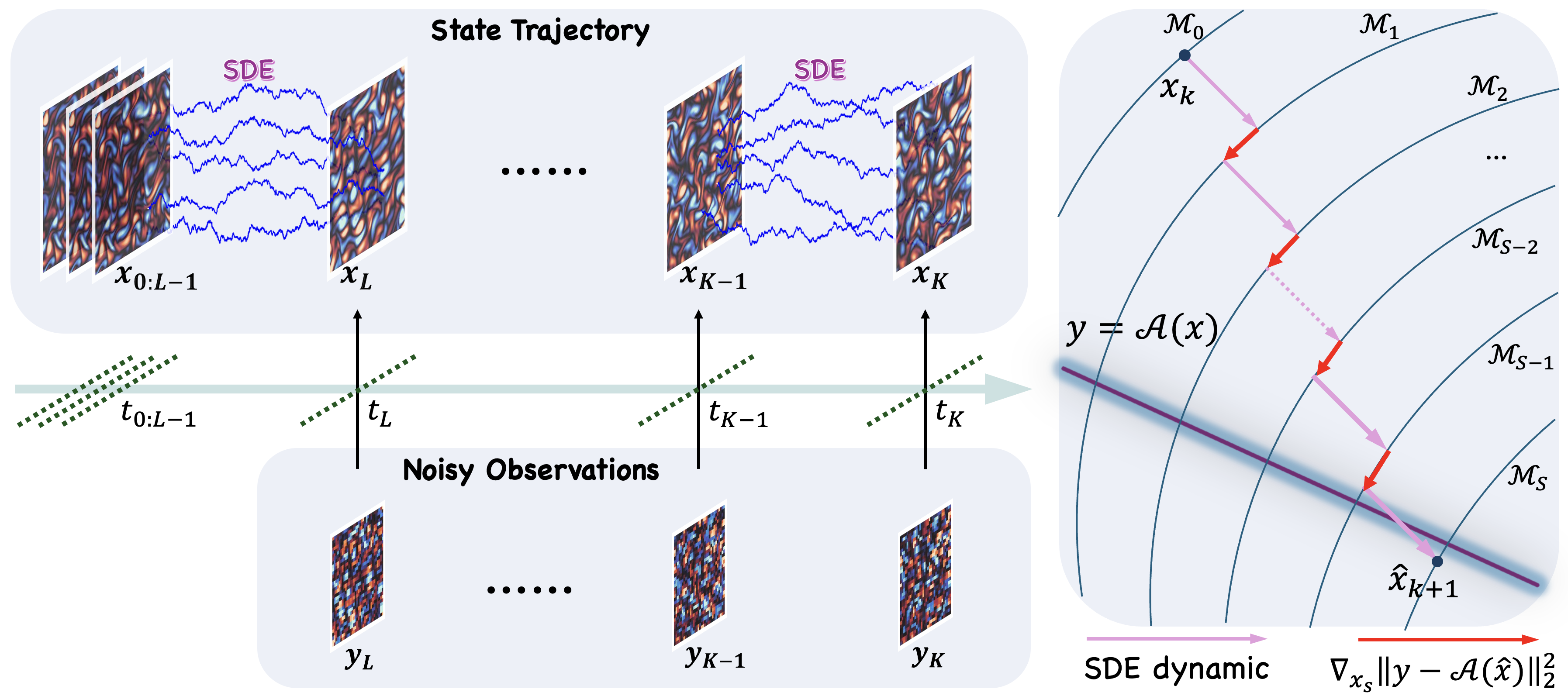}
\end{center}
\caption{\textbf{An overview of FlowDAS.}
We introduce a stochastic interpolant-based framework for data assimilation, named FlowDAS,
to estimate states $\bm x_{L:K}$ from the noisy (sparse or low-resolution) observations $\bm y_{L:K}$.
FlowDAS models the stochastic dynamics of the system with a flow-based stochastic differential equation (SDE)
and incorporates the observations to improve the prediction accuracy. 
On the right, we show a conceptual illustration of the geometry of the process to estimate $\hat{\x}_{k+1}$ from $\x_k$. $\calM_s$ denotes the generative manifold at interpolation step $s$. The gradient guidance $\nabla_{\x_s} \|\y - \calA (\hat{\x})\|_2^2$ enforces observation consistency during the generation process. 
}
\label{fig:overview}
\vspace{-0.2in}
\end{figure*}

Mathematically, a discrete-time stochastic dynamical system can be described by:
\begin{align}
    \xnext &= \Psi(\xk) + \bm \xi_k, 
    \label{eq:stochastic_dynamics_model} \\
     \bm y_{k+1} &= \cA(\xnext) + \bm \eta_{k+1}, 
    \label{eq:data_model}
\end{align}
where $\xk \in \mathbb{R}^D$ is the state vector at time step $k$,
$\Psi(\cdot)$ is the state transition map,
and $\bm \xi_k \in \mathbb{R}^D$ denotes the stochastic force.
The observations $\yk \in \mathbb{R}^M$ are related to the state
through the measurement map $\cA(\cdot)$,
with observation noise $\bm \eta_k \sim \cN(\mathbf 0, \gamma^2 \I_M)$. In many practical settings we have an initial window of known states
$\bm x_{0:L-1}$, after which observations $\bm y_{L:K}$ will only be noisy and partial.
DA seeks to infer the posterior of the state trajectory $\bm x_{L:K}$ given observations $\bm y_{L:K}$ and the initial states $\bm x_{0:L-1}$,
i.e., $p(\bm x_{L:K} \mid \bm y_{L:K}, \bm x_{0:L-1})$, as shown in \Cref{fig:overview}. Moreover, when the observations are absent, the task reduces to \emph{probabilistic forecasting} \cite{shysheya2024conditional}, where we predict $p(\bm x_{L:K}\mid\bm x_{0:L-1})$. 

Existing DA methods split into two categories. 
Model-driven methods---Kalman variants for quasi-linear/Gaussian systems \cite{evensen2003ensemble} and particle filters for fully nonlinear cases \cite{gordon1993novel}---need an accurate physical model and become costly or unstable in high dimensions \cite{bickel2008sharp}. 
Data-driven surrogates learn unknown or partially known dynamics directly from observations and embed them within Bayesian inversion frameworks. However, the intrinsic stochasticity of complex systems exposes limitations: neural operators such as FNO \cite{li2021fno} and Transolver \cite{wu2024transolver} produce only deterministic forecasts and thus cannot quantify uncertainty, while diffusion models \cite{qu2024deep,rozet2023score,shysheya2024conditional} struggle to learn physically faithful distributions due to the long Markov chain required to map Gaussian noise into realistic data. 

\paragraph{Our contributions.} We address the above limitations by building on recent flow-based \emph{stochastic interpolant} methods \cite{albergo2022building,albergo2023stochastic,pmlr-v235-chen24n}, which learn a short-step conditional transition $p(\x_{k+1}\!\mid\!\x_k)$ instead of a global noise-to-data map. Our main contributions are:
    
\begin{itemize}[leftmargin=*]
    \item \textbf{FlowDAS}: We introduce a generative data-assimilation framework that treats a learned stochastic interpolant as the forward model and assimilates new observations on the fly at inference time---without retraining or auxiliary filtering steps.

    \item \textbf{Efficient, interpretable learning}: By modeling the transition between adjacent states, FlowDAS trains faster and more stably than long‐horizon diffusion bridges, rolls out autoregressively, and provides clearer physical insight into one-step dynamics.

    \item \textbf{Extensive validation}: Across the Lorenz-63 system, incompressible Navier–Stokes super-resolution and sparse-sensor tasks, and large-scale SEVIR weather forecasting—where dynamics range from known to partially or fully unknown—FlowDAS consistently outperforms strong baselines in both accuracy and physical plausibility.
\end{itemize}

\vspace{-0.1in}
\section{Preliminaries}
\label{background}

\vspace{-0.1in}
\subsection{Related Work}
\vspace{-0.1in}
Data assimilation alternates between a \emph{forecasting} step---propagating the current state forward under a dynamical model---and an \emph{inverse} step that corrects this forecast with new observations. 
\vspace{-0.1in}
\paragraph{Model-driven methods} Traditional model‐driven approaches include the Kalman filter and its many variants \cite{asch2016data}. The Kalman filter provides an optimal minimum‐variance estimate when the dynamics are linear and all errors are Gaussian, but its performance degrades once these assumptions are not met. 
Also, there are some variants for nonlinear settings like the Extended Kalman Filter \cite{EKF1, EKF3}, applying the nonlinear transition to predict states and its Jacobian to the covariance, struggling in discontinuity situations and causing computational expense to calculate the Jacobian. 
The Unscented Kalman Filters \cite{UKF1} bypass the computation of Jacobian by propagating a set of sample points with the nonlinear transformations and deriving the mean and covariance, making it sensitive to sample point parameters and non-Gaussian problems. 
For fully nonlinear, non‐Gaussian problems, the bootstrap particle filter (BPF) \cite{gordon1993novel} approximates the posterior with a set of weighted particles that are advanced by the dynamics, reweighted by the observation likelihood, and resampled to prevent weight collapse. 
Although flexible, the BPF requires an exponential number of particles as the state dimension grows, making it impractical for large physical systems \cite{bickel2008sharp}. 
Besides, both the Kalman family and particle filters rely on having accurate knowledge of the underlying dynamical models, a requirement that is often unmet in real‐world applications.
\vspace{-0.1in}
\paragraph{Neural operator-based methods} Neural-operator solvers treat forecasting as a deterministic map from the current state to a single future state. The Fourier Neural Operator (FNO; \cite{li2021fno}) learns this map in Fourier space, giving fast and accurate predictions, while Transolver (\cite{wu2024transolver}) replaces Fourier layers with a physics-attention Transformer that can effectively capture complex physical correlations. Both models are deterministic and they are not designed for dynamics that are inherently stochastic. Recent work has sought to introduce stochasticity by fitting probabilistic surrogates---such as stochastic processes, graphical models, or stochastic Koopman operators \cite{wanner2022robust, zhao2022data}---but these efforts typically focus on low-dimensional settings or assume simple Gaussian statistics, limiting their applicability to high-dimensional, strongly nonlinear systems. 
Researchers have also adapted neural operator-based models for DA tasks \cite{chen2024fnp,xu2025fuxi,xiang2024adaf}. However, these adaptations are typically tailored to narrow use cases---such as arbitrary-resolution data assimilation---and have not demonstrated robust generalization across diverse real-world scenarios.
\vspace{-0.1in}
\paragraph{Diffusion model-based methods} Recent works have leveraged score-based methods, in particular diffusion models, for dynamical system modeling. For instance, the Conditional Diffusion Schrödinger Bridge (CDSB) framework extends diffusion models to conditional simulation and demonstrates strong performance on filtering in state-space models \cite{shi2022cdsb}. Score-based data assimilation (SDA) is a recent data-driven approach that employs score-based diffusion models to estimate state trajectories in dynamical systems \cite{rozet2023score}. It bypasses explicit physical modeling by learning the joint distribution of short state trajectory segments (e.g., $2k+1$ time steps) via a score network $\bm s_{\theta}(\bm x_{i-k:i+k})$. By integrating the score network with the observation model in a diffusion posterior sampling (DPS) framework \cite{chung2022diffusion}, SDA can generate entire state trajectories in zero-shot and non-autoregressive manners. However, SDA captures state transition implicitly from state concatenation, the learned dynamics of SDA may lack physical interpretability, leading to potential inaccuracies. 
And its posterior approximations may be less reliable in systems where the physical model is highly sensitive, as illustrated by the double well potential problem in \Cref{sec:double_well}. Additionally, diffusion models' non-autoregressive nature makes it hard to deal with long sequence rolling-out problems in high-dimension complex dynamics, as illustrated by the Navier–Stokes scenarios in \Cref{fig:ns_main}.

\vspace{-0.1in}
\subsection{Stochastic Interpolants}
\vspace{-0.1in}
Stochastic interpolants
\cite{albergo2022building, albergo2023stochastic, pmlr-v235-chen24n}
is a generative modeling framework
that unifies flow-based and diffusion-based models, which provides a smooth, controlled transition between arbitrary probability densities
over a finite time horizon.

Consider a stochastic process $\Xs$
defined over the interval \(s \in [0, 1]\),
which evolves from an initial state
\(
\X_0 \sim \pi(\X_0)\)
to the target state \( 
\X_1 \sim q(\X_1)\).
A stochastic interpolant can be described as:
\begin{equation}
    \Is = \alpha_s \X_0 + \beta_s \X_1 + \sigma_s \Ws,
    \label{eq:stocastic_interpolants}
\end{equation}
where \((\X_0, \X_1) \sim p(\X_0, \X_1)\).
$\Ws$
is a Wiener process
for $s \in [0, 1]$ introduced after $\X_0$ and $\X_1$ are sampled, ensuring that $\Ws$ is
independent of $\X_0$ and $\X_1$.
The time-varying coefficients 
\(\alpha_s, \beta_s, \sigma_s \in C^1([0, 1])\)
satisfy boundary conditions \(\alpha_0 = \beta_1 = 1\)
and \(\alpha_1 = \beta_0 = \sigma_1 = 0\), 
ensuring that \(\Izero = \X_0\) and \(\Ione = \X_1\),
thereby creating a smooth interpolation from \(\X_0\) to \(\X_1\).
Moreover, for all $(s, \X_0) \in [0,1] \times \mathbb{R}^D$,
$\Is \mid \X_0$ has the same distribution as
$\Xs$, which is the solution to the following SDE \cite{pmlr-v235-chen24n}:
\begin{equation}
    \der \Xs = \bs(\Xs, \X_0) \der s + \sigma_s \der \Ws,
    \label{eq:sde}
\end{equation}
where the drift term \(\bs(\X, \X_0)\) is optimized by minimizing
the cost function:
\begin{equation}
    \cL_b (\hat{\blb}_s) = \int_0^1 \mathbb{E} \left[\| \hat{\blb}_s(\Is, \X_0) - \Rs \|^2 \right] \der s.
    \label{eq:objective_bs}
\end{equation}
The ``velocity'' of the interpolant path, \(\Rs\), is given by $\Rs = \dot{\alpha}_s \X_0 + \dot{\beta}_s \X_1 + \dot{\sigma}_s \Ws$. Furthermore, the drift term $\bs(\Xs, \X_0)$ is related to the score function
$\nabla \log p(\Xs \mid \X_0)$: 
\begin{equation}
\begin{aligned}
    \bs(\Xs, \X_0) &= \frac{\cs(\Xs, \X_0)}{\beta_s}
    + \frac{\nabla \log p(\Xs \mid \X_0)}{\As \beta_s},
    \label{eq:score}
\end{aligned}
\end{equation}
where \As and \(\cs(\Xs, \X_0)\) are defined to control the score-based dynamics (\Cref{sec:relate_drift_and_score}). 

Building on this framework, stochastic interpolants can estimate the transition $p(\bm x_{k+1}\mid\bm x_k)$ by evolving a latent path $\bm X_s$ smoothly from $s=0$ (current state) to $s=1$ (next state) in dynamical systems. It provides a probabilistic yet compact surrogate that is more aligned with the true step-to-step evolution of complex dynamical systems. While stochastic interpolants capture complex system dynamics, they does not yet enforce consistency with measurements.

\vspace{-0.1in}
\section{Method}
\label{sec:method}
\vspace{-0.1in}
\subsection{Stochastic Interpolants for Data Assimilation}
\vspace{-0.1in}
Using stochastic interpolants as the engine of data assimilation is attractive but \emph{not} a plug-and-play replacement for existing forward models: it raises \emph{three fundamental challenges} that FlowDAS must overcome. Below we formulate each challenge and describe the corresponding FlowDAS solution.

\vspace{-0.1in}
\paragraph{Challenge I: Observation-consistent state generation} 
Stochastic interpolants approximate the state transition
\(p(\xnext \mid \xk) \)
by interpolating between the state variables \xk and \xnext
using the SDE defined in \Cref{eq:sde} and \Cref{eq:score} with boundary conditions $\X_0, \X_1 = \xk, \xnext$. Moreover, DA forward surrogate must respect the observation $\y_{k+1}$ while drawing the future state $\x_{k+1}$.
For stochastic interpolants, this requires an \emph{observation-conditioned} SDE. FlowDAS augments the original drift $\bs(\Xs,\X_0)$ via Bayes' rule (\Cref{sec:conditional_drift}):
\begin{align}
   \bs(\Xs, \y,\X_0)=\bs(\Xs, \X_0) + \frac{\nabla \log p(\y \mid \Xs, \X_0)}{\As \beta_s}.
    \label{eq:bs_g_bayes}
\end{align}

\vspace{-0.1in}
\paragraph{Challenge II: Estimating the conditional score} 
The term \(\nabla \log p(\y \mid \Xs, \X_0)\) captures the observation information, however, it is intractable because the observation model only directly links
$\y = \y_{k+1}$ and $\X_1 = \xnext$. FlowDAS estimates it by Monte-Carlo marginalization, i.e., integrating with respect to $\X_1$, which can be approximated by $J$ Monte Carlo samples $\X_1^{(j)} \sim p(\X_1 \mid \Xs, \X_0)$:
\begin{equation}
\begin{aligned}
    \nabla \log p(\y \mid \Xs, \X_0) = \frac{\nabla \,\mathbb{E}_{\X_1  \sim p(\X_1 \mid \Xs, \X_0)}[p(\y|\X_1)]}{\mathbb{E}_{\X_1  \sim p(\X_1 \mid \Xs, \X_0)}[p(\y|\X_1)]} \approx \sum_{j=1}^{J} w_j \nabla \, \log p(\y \mid \X_1^{(j)}),
\end{aligned}
\label{eq:jensen_estimation_mc}
\end{equation}
where we apply
a softmax function to the $J$ scalars
$\{\log p(\y \mid \X_1^{(j)})\}_{j=1}^{J}$
to compute the sample weights
$w_j = p(\y \mid \X_1^{(j)})/\sum_{j=1}^{J}  p(\y \mid \X_1^{(j)})$. We leave detailed derivation in \Cref{sec:estimate_loglikelihood}.

\vspace{-0.1in}
\paragraph{Challenge III: Efficient Monte Carlo sampling} 
Accurate sampling from \( p(\X_1 \mid \Xs, \X_0) \)
requires solving the SDE in \Cref{eq:sde},
which can be computationally intensive. FlowDAS accelerates this step with low-order stochastic integrators such as the first-order Milstein method \cite[p.~317]{suli2003introduction}:
\begin{equation}
    \hat {\X}_1 = \X_0 + \bs(\Xs, \X_0)(1-s)+\int_0^1 \sigma_s \der \Ws
    \label{eq:1st_integration}
\end{equation}
and the second-order stochastic Runge-Kutta method \cite[p.~324]{suli2003introduction}: 
\begin{equation}
    \hat {\X}_1^{\prime} = \X_0 + \frac{\bs(\Xs, \X_0)+\blb_1(\hat {\X}_1, \X_0)}{2}
    (1-s) 
    +\int_0^1 \sigma_s \der \Ws.
    \label{eq:2nd_integration}
\end{equation}
Both approximations introduce slight numerical bias (i.e., $\cO((1 - s)^2)$ and $\cO((1 - s)^3)$, respectively)
but significantly accelerate sampling speed. 
\Cref{sub : The Method For Posterior Estimation}
further compares these approximations.

\vspace{-0.1in}
\paragraph{Advantages of FlowDAS.} Compared with other data-driven surrogates, FlowDAS offers two key advantages. 
\begin{itemize}[leftmargin=*]
    \item \textbf{Observation-consistent probabilistic forecasts.} Neural operators such as FNO and Transolver learn a deterministic map $\x_k \!\mapsto\! \x_{k+1}$; they yield a single forecast and require a separate optimisation step to reconcile that forecast with the observation $\y_{k+1}$. FlowDAS learns the full conditional distribution $p(\bm x_{k+1}\mid \bm x_k,\bm y_{k+1})$. It therefore generates an ensemble of forecasts that are already consistent with the incoming measurement, eliminating the need for any post-hoc update.
    
    \item \textbf{Local and physics-aligned transport.} FlowDAS directly learns a \emph{local} bridge between adjacent states, so the diffusion process spans a short distance in state space. Training therefore remains numerically stable and provides clearer physical insight into one-step dynamics. By contrast, score-based diffusion models construct a global path from Gaussian noise to data; the network must master a far longer transformation and learn the state transition dynamics through channel correlations, which often obscures physical interpretability and can degrade accuracy. 
\end{itemize}

\vspace{-0.1in}
\subsection{Implementation Details}

\vspace{-0.1in}
\paragraph{Training} We train a user-defined neural network as the drift model \( \bs(\Xs, \X_0) \), which outputs the velocity given the current interpolant \( \Xs \) and conditioned on the initial state \( \X_0 \). The network architecture of the drift model may vary across tasks, for example, an Multi-Layer Perceptron (MLP) is adopted for the low-dimensional Lorenz-63 system in \Cref{sec:lorenz}, a U-Net for high-dimensional fluid system governed by incompressible Navier-Stokes equations in \Cref{sec:NS}, and a FNO for the weather forecasting task in \Cref{sec:sevir}. 
The model is optimized using collected trajectories \( \x_{0:K} \), minimizing an empirical loss between predicted and target velocities over sampled interpolation times $s \in [0, 1]$. The discrepancy metric (e.g., $\ell_1$ or squared $\ell_2$ norm, etc.) is also task-dependent.

\vspace{-0.1in}
\paragraph{Inference} Given a trained drift model \( \hat{\blb}_s \), inference proceeds in an autoregressive manner. Starting from a known state \( \xk \), we define a discretized interpolation interval $s_0 = 0 < s_1 < \dots < s_N = 1$. At each step, we use multiple posterior samples \( \hat{\X}_1^{(j)} \) to update \( \X_{s_n} \) by enforcing consistency with observed measurements \( \y_{k+1} \). The updated state at $s_N$ becomes the initial state for the next time step, i.e., \( \X_{s_0}^{k+1} = \X_{s_N}^{k} \). This process is repeated to generate the full trajectory. \Cref{sec:implementation} provides further details on our implementation and the algorithm.

\vspace{-0.1in}
\paragraph{Conditioning} While the original stochastic interpolants framework which only models the \(p(\xnext \mid \xk) \), we take in a sequence of previous states to achieve the transition from  \(p(\xk \mid \x_{k-1}...\x_{k-l}) \) to \(p(\xnext \mid \xk, \x_{k-1}...\x_{k-l}) \) inspired by \cite{voleti2022mcvd, Yu_2023_CVPR, Zhang_2024_CVPR, gao2023prediff}, thus achieving probabilistic prediction conditioned on several previous states. 
This straightforward extension allows FlowDAS to handle non-Markovian dynamics and empirically leads to markedly improved performance on weather-forecasting task (\Cref{sec:sevir}).

\vspace{-0.1in}
\section{Experiments and Results}
\label{sec:results}

\vspace{-0.1in}

This section evaluates our proposed framework, FlowDAS, on a range of
low- and high-dimensional stochastic dynamical systems, including low-dimensional problems with high-order observation models, such as the double-well potential (\Cref{sec:double-well}) and the chaotic Lorenz 1963 system, as well as high-dimensional tasks involving the incompressible Navier-Stokes equations and a realistic problem: Particle Image Velocimetry (PIV). 
Additionally, we demonstrate its applicability on a real-world large-scale weather forecasting task, where governing dynamics are not available. 
These results underscore the versatility and robustness of FlowDAS.

\vspace{-0.1in}
\subsection{Lorenz 1963}
\label{sec:lorenz}

\vspace{-0.1in}
In this experiment, we evaluate the performance of FlowDAS using the Lorenz 1963 system,
a simplied mathematical model for atmospheric convection that is widely studied in the DA community
\cite{Lorenz1963DeterministicNF, rozet2023score, bao2023scorebasednonlinearfilterdata}. 
The state vector of the Lorenz system, $\x = (a, b, c) \in \mathbb{R}^3$,
evolves according to the following
nonlinear stochastic ordinary differential equations (ODEs):
\begin{equation}
\begin{aligned}
\Frac{\der a}{\der t} &= \mu (b - a) + \xi_1, \\
\Frac{\der b}{\der t} &= a(\rho - c) - b + \xi_2, \\
\Frac{\der c}{\der t} &= ab - \tau c + \xi_3,
\end{aligned}
\label{eq:lorenz transition}
\end{equation}
where $\mu = 10$, $\rho = 28$, and $\tau = \Frac{8}{3}$ define the ODE parameters, and $\bm \xi = (\xi_1, \xi_2, \xi_3)\in \mathbb{R}^3$  is the process noise, with each component having a standard deviation $\sigma = 0.25$.
This chaotic system poses a significant challenge for numerical methods, so we use the fourth-order Runge-Kutta (RK4) method \cite{10.1063/5.0192085} to simulate its state transition (see \Cref{sec:supp in lorenz} for details).

We observe only the arctangent-transformed value of the first state component $a$, so the observation model of the system is defined as 
\begin{equation}
y = \cA(\x) + \bm \eta = \arctan(a) + \eta,
\end{equation}
wehre $\eta$ is the observation noise with a standard deviation $\gamma = 0.25$.

\vspace{-0.1in}
\paragraph{Dataset and experiments}
We generate 1,024 independent trajectories, each containing 1,024 states, and split the data into training (80\%), validation (10\%), and evaluation (10\%) sets. Initial states are sampled from the statistically stationary regime of the Lorenz system, with additional data generation details provided in \Cref{sec:supp in lorenz}. During inference, we independently estimate 64 trajectories over 15 time steps using FlowDAS and the baseline methods. A total of $L=1$ previous state is conditioned for each autoregressive generation. For this low-dimensional problem, we use a fully connected neural network to approximate the drift term in stochastic interpolants; the network architecture is also described in the same appendix.

\vspace{-0.1in}
\paragraph{Baselines and metrics}
We compare our method against two baselines:
the SDA solver with a fixed window size of 2 and the classic BPF \cite{gordon1993novel}. \Cref{sec:supp in lorenz} details the score network architecture for SDA and particle density settings for BPF. 

We evaluate the performance of FlowDAS and baselines using four metrics:
the expectation of log-prior
$\textstyle \mathbb{E} \left[ \log p(\hat{\x}_{2:K} \mid \hat{\x}_1) \right]$;
the expectation of log data likelihood
$\textstyle \mathbb{E} \left[ \log p(\y_{1:K} \mid \hat{\x}_{1:K}) \right]$;
the Wasserstein distance \cite{Villani2009}
$W_1(\cdot, \cdot)$ between the true trajectory $\x_{1:K}$
and the estimated trajectory $\hat{\x}_{1:K}$;
and the RMSE between the true and estimated states.

\begin{figure*}[t]
\begin{center}
\centerline{\includegraphics[width=0.9\textwidth]{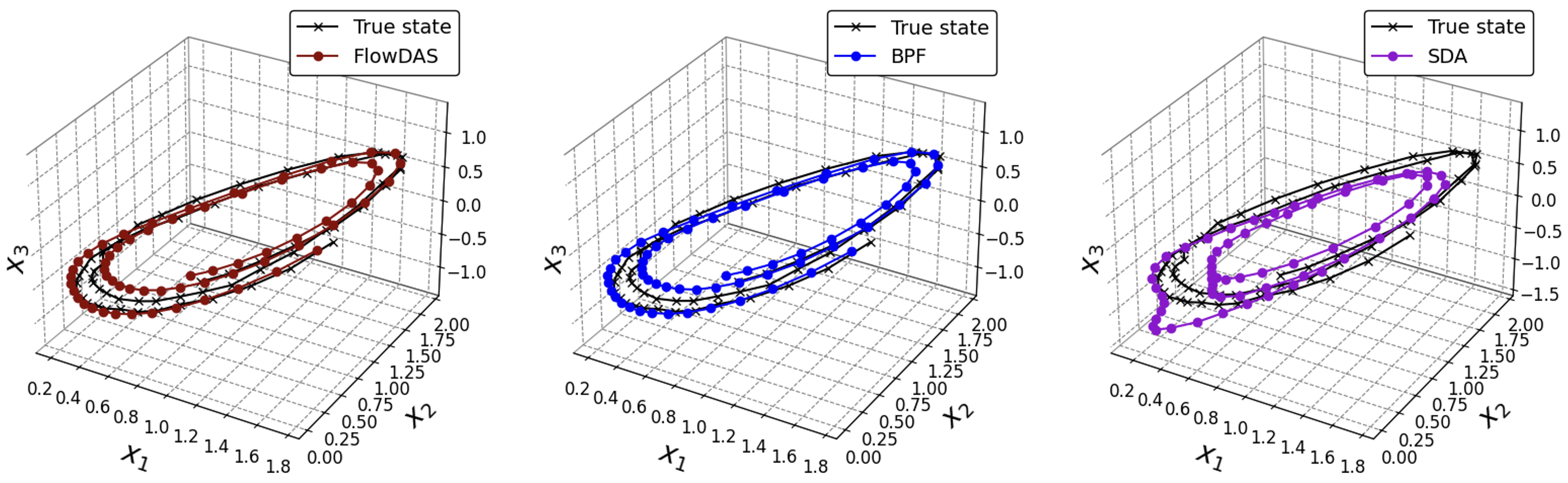}}
\caption{
\textbf{Data assimilation of Lorenz 1963 system.}
FlowDAS achieved results comparable to the state-of-the-art model-based BPF method,
significantly outperforming the data-driven SDA method
in recovering the underlying dynamics of this chaotic system.
This highlights the efficiency and robustness of FlowDAS
in capturing complex, nonlinear dynamics while maintaining accuracy and stability.
The variables $x_1$, $x_2$ and $x_3$ correspond to $a$, $b$ and $c$
in the ODEs of the Lorenz system \Cref{eq:lorenz transition}, respectively.
}
\label{fig:lorenz_trajectory}
\end{center}
\vspace{-0.3in}
\end{figure*}

\begin{table}[!h]
\begin{center}
\begin{sc}
\begin{tabular}{clcc|ccr}
\toprule
& & FlowDAS & SDA & BPF \\
\midrule
$ \log p(\hat{\x}_{2:K} \mid \hat{\x}_1)$  & $\uparrow$ & 17.29 & -332.7& \textbf{17.88}  \\
$\log p(\y_{1:K} \mid \hat{\x}_{1:K})$ & 
$\uparrow$ & \textbf{-0.228} & -6.112 & -1.572 \\
$W_1(\x_{1:K}, \hat{\x}_{1:K})$ & $\downarrow$ & \textbf{0.106} & 0.528&  0.812  \\
RMSE$(\x_{1:K}, \hat{\x}_{1:K})$  & 
$\downarrow$ & \textbf{0.202}      &  1.114    &     0.270     \\
\bottomrule
\end{tabular}
\end{sc}
\end{center}
\caption{\textbf{Data assimilation of Lorenz 1963 system.} This table summarizes the performance of FlowDAS, SDA, and BPF on the Lorenz 1963 experiment over 15 time steps. FlowDAS outperforms SDA across all evaluation metrics and is competitive with BPF, despite BPF utilizing the true transition equations, which are unknown to FlowDAS and SDA. The best results for each metric are highlighted in \textbf{bold}.
}
\label{lorenz eva}
\vspace{-0.1in}
\end{table}

\vspace{-0.1in}
\paragraph{Results}
As shown in \Cref{lorenz eva} and \Cref{fig:lorenz_trajectory}, FlowDAS outperforms SDA across all metrics. FlowDAS is only slightly less effective than BPF in the expected log-prior, as BPF directly incorporates the true system dynamics into its state estimation.

The success of FlowDAS is primarily due to its accurate mapping from current to future states ($\xk \rightarrow \xnext$). Despite lacking explicit transition equations, FlowDAS effectively captures the system dynamics through stochastic interpolants, enabling a closer approximation of state trajectories compared to SDA, which models joint distributions across sequential states using diffusion models. Additionally, stochastic interpolants allow FlowDAS to produce accurate state estimates while managing inherent variability, avoiding over-concentration on high-probability regions, and effectively dealing with rare events. This advantage is further illustrated in the double well potential experiment (\Cref{sec:double-well}) where FlowDAS outperforms BPF, because BPF tends to be trapped by high-probability point estimates.

\vspace{-0.1in}
\subsection{Incompressible Navier-Stokes Flow}
\label{sec:NS}

\begin{figure*}[t]
\begin{center}
\centerline{\includegraphics[width=0.9\textwidth]{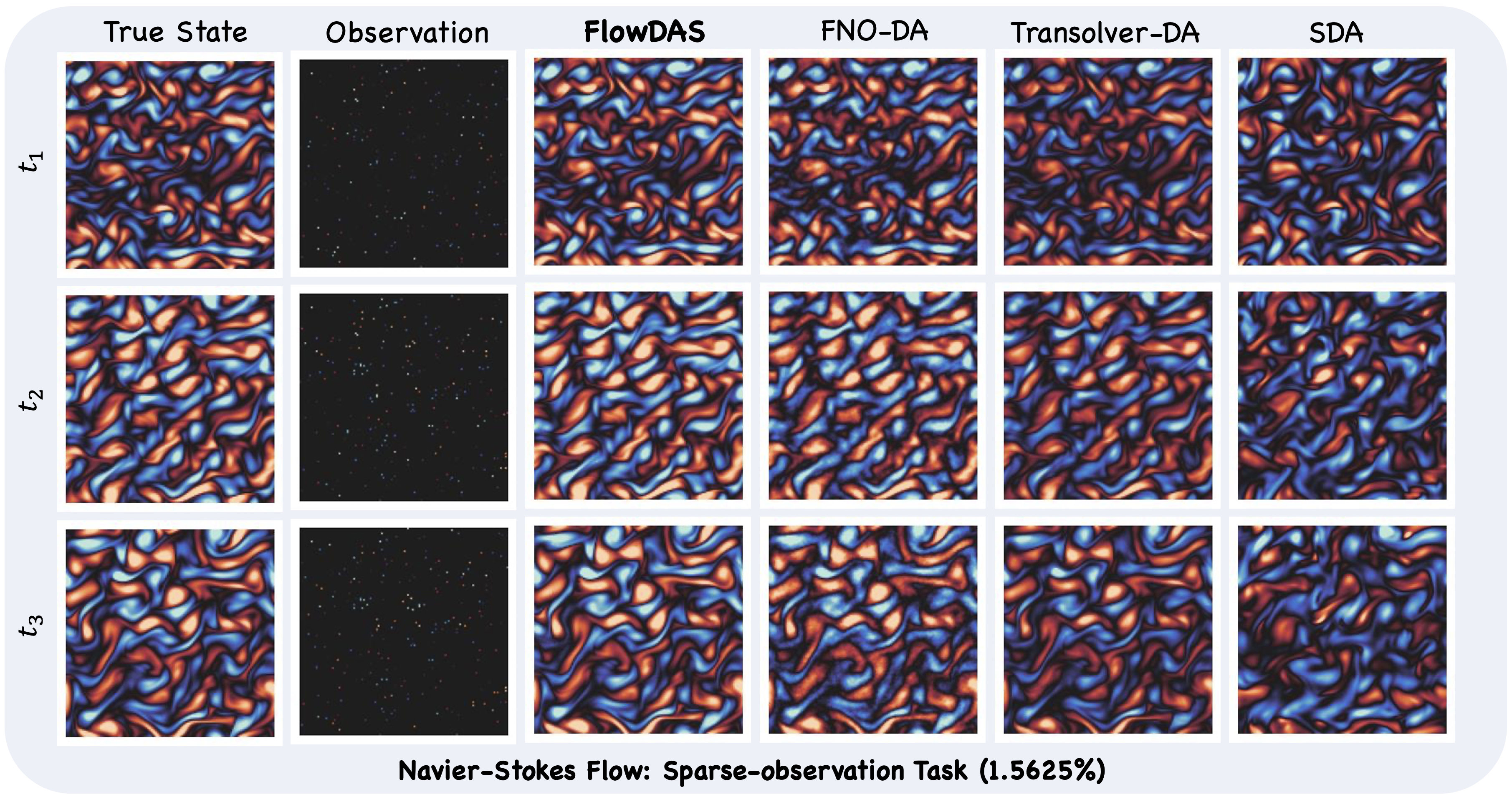}}
\caption{\textbf{Data assimilation of incompressible Navier-Stokes flow. }
The positive values (red) of the state, i.e., vorticity field, indicate clockwise rotation and negative values (blue) indicate counter-clockwise rotation. FlowDAS achieved results with more accurate details and higher accuracy than all baselines, showing the efficiency of FlowDAS in tackling DA tasks with highly non-linear complex systems. Additionally, FlowDAS is also better at recovering high-frequency information, evidenced by the spectral analysis in \Cref{fig:spectral_4}.}
\label{fig:ns_main}
\end{center}
\vspace{-0.3in}
\end{figure*}

This section considers a high-dimensional dynamical system:
incompressible fluid flow governed by the 2D Navier-Stokes (NS) equations with random forcing on the torus $\mathbb{T}^2 = [0, 2\pi]^2$. The state transition, $\Psi$, is described using the stream function formulation, 
\begin{equation}
    \der \bm \omega + \blv \cdot \nabla \bm \omega \, \der t
    = \nu \Delta \bm \omega \, \der t - \alpha \bm \omega \, \der t + \varepsilon \der \bm \xi,
\label{eq:ns equation}
\end{equation}
where $\bm \omega$ represents the vorticity field, the state variable in this fluid dynamics system ($\x = \bm \omega$). 
The velocity $\blv = \nabla^\perp \psi = (-\partial_y \psi, \partial_x \psi)$ is expressed in terms of the stream function $\psi (x, y)$, which satisfies $-\Delta \psi = \bm \omega$. The term $d\bm \xi$ represents white-in-time random forcing acting on a few Fourier modes, with parameters $\nu, \alpha, \varepsilon > 0$ specified in \Cref{subsub: problem settings of ns}.

The observation operator \cA linearly downsamples or selects partial pixels from the simulated vorticity fields ($\bm \omega$), 
\begin{equation}
    \y = \cA (\bm \omega) + \bm \eta,
\end{equation}
where the observation noise $\bm \eta$ has a standard deviation of $\gamma = 0.05$. 

\vspace{-0.1in}
\paragraph{Dataset and experiments}
In this experiment, system dynamics are simulated by solving \Cref{eq:ns equation} using a pseudo-spectral method \cite{peyret2002spectral} with a resolution of $256^2$ and a timestep $\Delta t = 10^{-4}$. We simulate 200 flow conditions over $t \in [0, 100]$, saving snapshots of fluid vorticity field ($\bm \omega = \nabla \times v$) at the second half of each trajectory ($t \in [50, 100]$) at intervals of $\Delta t = 0.5$ with a reduced resolution of $128^2$. The data are divided into training (80\%), validation (10\%), and evaluation (10\%) sets. 

We conduct experiments across different observation resolutions ($32^2$, $16^2$) and observation sparsity levels (5\%, 1.5625\%). For the super-resolution task, the goal is to reconstruct high-resolution vorticity data ($128^2$) from low-resolution observations. In the inpainting task, only 5\% or 1.5625\% of pixel values are retained, with the rest set to zero, and we attempt to recover the complete vorticity field. The model is evaluated on four unseen datasets (2$\times$ super-resolution, 2$\times$ inpainting), with 64 samples for each configuration. A total of $L=10$ previous states are conditioned for each autoregressive generation.

\vspace{-0.1in}
\paragraph{Baselines and metrics} 
We benchmark FlowDAS against three data-driven baselines.
First, we implement SDA solver with a fixed eleven-step window (10 previous states as conditions).
Second, we adapt the deterministic neural operators FNO and Transolver to the DA setting, namely `FNO-DA' and `Transolver-DA'. Each network is trained to minimise RMSE between adjacent states; at inference time the raw prediction $\x'_{k+1}=f_\theta(\x_k)$ is reconciled with the observation $\y_{k+1}$ by solving
$
\hat{\x}_{k+1}
=\argmin_{\x}\,
\alpha_1\|\x - \x'_{k+1}\|_2^2+
\alpha_2\|\y_{k+1}-\calA(\x)\|_2^2,
\label{eq:extend_baseline_opt}
$
an optimization step analogous to a Kalman update, where $\alpha_1$ and $\alpha_2$ are weighting coefficients. The $\alpha_1$ and $\alpha_2$ are chosen by grid searching, and we report the best results. BPF is not included in our testing, as its particle requirements grow exponentially with system dimensions, making it impractical for high-dimensional fluid dynamics systems. Additional details on the model architecture and training for baselines and FlowDAS are provided in \Cref{subsub sec:details about the ns neural network}.

\noindent We evaluate performance using the RMSE between the predicted and ground-truth vorticity fields. Additionally, we assess the reconstruction of the kinetic energy spectrum to determine whether the physical characteristics of the fluid are accurately preserved. \Cref{subsubsec: detail of kinetic energy spectrum} and \Cref{fig:spectral_4} provide the definitions and results for the kinetic energy spectrum metric.

\begin{table}[t]
\begin{center}
\begin{sc}
\setlength{\tabcolsep}{3pt} 
\begin{tabular}{lcccr}
\toprule
  & $32^2 \rightarrow 128^2$ & $16^2 \rightarrow 128^2 $ & $5\%$ &$1.5625\%$\\
\midrule
FlowDAS &  \textbf{0.038} & \textbf{0.067} & \textbf{0.071} & \textbf{0.123}\\
FNO-DA & 0.158 & 0.166 & 0.165 & 0.183 \\
Transolver-DA & 0.159 & 0.176 & 0.161 & 0.180 \\
SDA & 0.073 & 0.133 & 0.251 & 0.258 \\

\bottomrule
\end{tabular}
\end{sc}
\end{center}
\vskip 0in
\caption{\textbf{RMSE of FlowDAS and baselines on incompressible Naiver-Stokes super-resolution and sparse observation tasks.} Results are reported for various settings:
super-resolution from $32^2$ and $16^2$ to $128^2$
and sparse observation recovery at 5\% and 1.5625\% observed pixels.
The best results are highlighted in $\textbf{bold}.$} 
\label{tab:NS}
\vspace{-0.2in}
\end{table}

\vspace{-0.1in}
\paragraph{Results}
\Cref{fig:ns_main} compares FlowDAS with baselines under the 1.5625\% sparse-observation setting. See \Cref{sec:ns_rest} for additional results. Quantitative metrics, including RMSE and kinetic energy spectrum comparisons, are provided in \Cref{tab:NS} and \Cref{fig:spectral_4}. Our method consistently outperforms all baselines in terms of reconstruction accuracy, capturing high-frequency information with greater precision. This advantage is further validated by the kinetic energy spectrum in \Cref{fig:spectral_4}. The RMSE scores in \Cref{tab:NS} further highlight the effectiveness of FlowDAS in accurately estimating the underlying fluid dynamics from observational data.

\vspace{-0.1in}
\paragraph{Particle Image Velocimetry} 
To demonstrate that our framework generalizes beyond the synthetic incompressible NS setting, we also apply FlowDAS to Particle Image Velocimetry, a widely used optical technique that infers sparse planar velocity vectors by tracking tracer particles in sequential images. In this scenario the DA problem is to reconstruct the full vorticity (or velocity) field from noisy, sparsely sampled velocities. We generate synthetic PIV frames from the same NS simulations, extract particle displacements with a standard PIV pipeline, and feed the resulting observations into the pretrained FlowDAS model. A comparison with baseline methods confirms that FlowDAS maintains its performance advantage in this practical setting. The full experimental protocol, quantitative metrics, and qualitative results are reported in \Cref{sec:supp_on_PIV}.


\begin{figure}[t]
\centering
\includegraphics[width=0.9\linewidth]{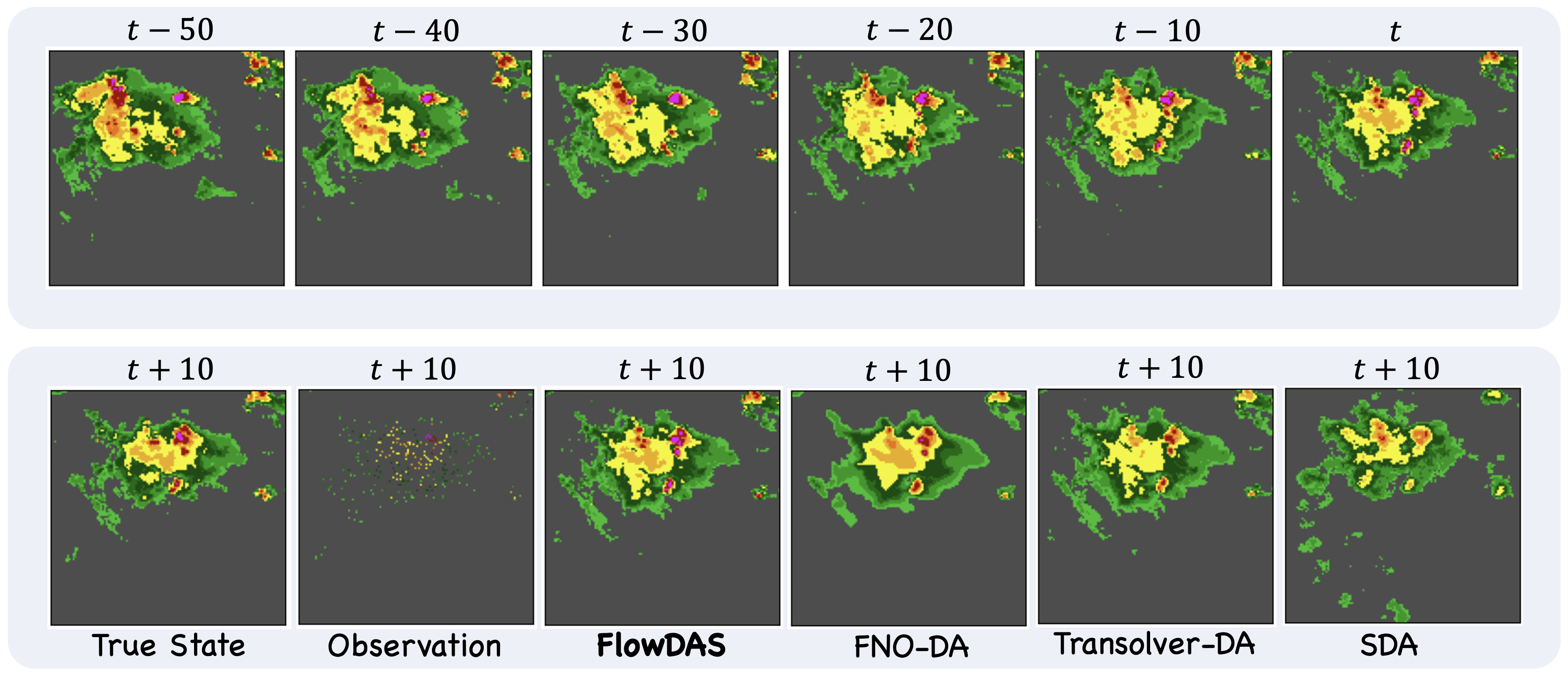}
\caption{\textbf{Data assimilation of weather forecasting on SEVIR Vertical Integrated Liquid dataset under sparse observations.} All DA models take previous six states (displayed in the first row; $t\!-\!50$ min to $t$ min) as conditions and estimate the future state at $t+10$ min. FlowDAS preserves storm cores and spatial texture better than score-based diffusion models and neural operator-based methods.}
\label{fig:sevir_vis}
\vspace{-0.2in}
\end{figure}

\vspace{-0.1in}
\subsection{Weather Forecasting}
\label{sec:sevir}
This experiment evaluates FlowDAS on a large‐scale, real‐world weather–forecasting problem using the Storm EVent Imagery and Radar (SEVIR) dataset \cite{veillette2020sevir}. SEVIR provides multi‐modal observations of severe convective storms across the continental United States.  We focus on the \emph{Vertically Integrated Liquid} (VIL) product, a 2‐D proxy for precipitation intensity. Each sample is a $128\times128$ grid covering $384\text{ km}\times384\text{ km}$ at 2 km resolution and recorded every 10 min for four hours.  Similar to \cite{gao2023prediff}, six consecutive VIL frames ($t\!-\!50$ min to $t$ min) constitute the input, and the task is to predict the next frame at $t+10$ min. In our DA settings, the state variable is the VIL field $\x$, and the observation is aquired by randomly sampling 10\% of grid cells to emulate sparse radar coverage. A total of $L=6$ previous states are conditioned for next-frame generation. We split the dataset into 80\%/10\%/10\% for training, validation, and testing.

\begin{table}[t]
\begin{center}
\begin{sc}
\begin{tabular}{lccc}
\toprule
\textbf{Method} & \textbf{RMSE} $\downarrow$ & \textbf{CSI($\tau_{20}$) (0.3)} $\uparrow$ & \textbf{CSI($\tau_{40}$) (0.5)} $\uparrow$\\
\midrule
Transolver-DA       & 0.062$\pm$0.001 & 0.663$\pm$0.001 & 0.499$\pm$0.002 \\
FNO-DA              & 0.064$\pm$0.001 & 0.641$\pm$0.001 & 0.493$\pm$0.002 \\
SDA  & 0.071$\pm$0.007 & 0.549$\pm$0.033 & 0.387$\pm$0.065 \\
FlowDAS          & \textbf{0.053$\pm$0.004} & \textbf{0.746$\pm$0.022} & \textbf{0.614$\pm$0.044} \\
\bottomrule
\end{tabular}
\end{sc}
\end{center}
\caption{Comparison of FlowDAS and other baselines on RMSE and Critical Success Index (CSI) of the weather forecasting task. All metrics are averaged over multiple runs with standard deviations.}
\label{tab:sevir}
\end{table}

\paragraph{Baselines and metrics} We compare FlowDAS with the SDA solver (fixed window size of 7 with 6 previous states as conditions). Neural-operator baselines, e.g., FNO and Transolver, are similarly obtained as Navier-Stokes experiments. Both models takes a concatenation (along channel dimension) of states as inputs and output the state at target time step. Performance is measured by RMSE and the Critical Success Index (CSI) at thresholds of $\tau_{20}$ and $\tau_{40}$ dBZ, which are standard verification metrics for precipitation nowcasting. \Cref{sec:supp_on_SEVIR} provides more experimental details.

\paragraph{Inplementation Details} Different from \cite{pmlr-v235-chen24n}, we adopted FNO
\cite{li2021fno} rather than U-Net as the backbone of the stochastic interpolants framework. We provide the comparison between the FNO-based and the U-Net-based FlowDAS in the \Cref{sec:supp_on_SEVIR}, and found that the FNO-based FlowDAS demonstrates better performance.

\paragraph{Results} FlowDAS achieves lower RMSE and higher CSI at both thresholds as shown in \Cref{tab:sevir}, indicating more accurate intensity estimates and better hit rates for heavy precipitation. Qualitative comparisons in \Cref{fig:sevir_vis} and \Cref{fig:sevir1,fig:sevir2,fig:sevir3} show that FlowDAS reconstructs coherent \emph{precipitation structures} and \emph{peak intensities} that other baselines either smooth out or mis-localize. These results demonstrate that the stochastic‐interpolant surrogate scales to high-resolution, real meteorological data, even when underlying governed PDEs are unknown, and retains its advantage over diffusion models and neural operators baselines.

\section{Conclusion and Future Work}
In this work, we introduced \textbf{FlowDAS}, a stochastic interpolant-based data assimilation framework designed to address the challenges of high-dimensional, nonlinear dynamical systems. By leveraging stochastic interpolants, FlowDAS effectively integrates complex transition dynamics with observational data, enabling accurate state estimation without relying on explicit physical simulations. Through experiments on both low- and high-dimensional systems---including the Lorenz 1963 system, incompressible Navier-Stokes flow, and weather forecasting---FlowDAS demonstrated strong performance in recovering accurate state variables from sparse, noisy observations. These results highlight FlowDAS as a robust alternative to traditional model-driven methods (e.g., particle filters) and data-driven approaches (e.g., score-based diffusion models and neural operators) for data assimilation, offering improved accuracy, efficiency, and adaptability. Future work will focus on scaling the framework to more complex real-world systems~\cite{pmlr-v235-huang24h,Seabra2024AIED}, exploring its generalizability \cite{yasuda2023spatio}, and further optimizing its performance under Sim2Real setting \cite{10.1063/5.0209339}.





\bibliographystyle{abbrvnat}
\bibliography{reference}

\addcontentsline{toc}{section}{Appendix}

\addtocontents{toc}{\protect\setcounter{tocdepth}{0}}  

\newpage
\appendix

\renewcommand{\thefigure}{S.\arabic{figure}}
\renewcommand{\theequation}{S.\arabic{equation}}
\renewcommand{\thetable}{S.\arabic{table}}

\appendices

\section{Mathematical Derivation}
\label{sec:maths}

\subsection{Conditional Drift}
\label{sec:conditional_drift}
Stochastic interpolants approximate the state transition
\(p(\xnext \mid \xk) \)
by interpolating between the state variables \xk and \xnext
using the SDE defined in \Cref{eq:sde} and \Cref{eq:score} with boundary conditions $\X_0, \X_1 = \xk, \xnext$. The drift term $\bs(\Xs, \X_0)$ is related to the score function
$\nabla \log p(\Xs \mid \X_0)$
of state transition distribution (\Cref{sec:relate_drift_and_score}): 
\begin{equation}
\begin{aligned}
    \bs(\Xs, \X_0) &= \frac{c_s(\Xs, \X_0)}{\beta_s}
    + \frac{\nabla \log p(\Xs \mid \X_0)}{\As \beta_s}.
    \label{eq:original stochastic interpolant and score}
\end{aligned}
\end{equation}
To generate observation-consistent states,
we modify the SDE to a process conditioned on observational data, where the drift term $\bs(\Xs, \y,\X_0)$
incorporates observation information via Bayes' rule: 
\begin{align}
   \bs(\Xs, \y,\X_0) =& \frac{c_s(\Xs , \X_0)}{\beta_s}
    + \frac{\nabla \log p(\Xs \mid \y, \X_0)}{{\As \beta_s}}  \nonumber \\
    =&\frac{c_s(\Xs, \X_0)}{\beta_s}+\frac{\nabla \log p(\y \mid \Xs, \X_0)
    + \nabla \log p(\Xs \mid \X_0)}{\As \beta_s} \nonumber \\
    =&\bs(\Xs, \X_0) + \frac{\nabla \log p(\y \mid \Xs, \X_0)}
   {\As \beta_s}.
    \label{eq:bs_g_bayes_append}
\end{align}

\subsection{Relating Drift and Score}
\label{sec:relate_drift_and_score}
By extending Stein's formula \cite{albergo2023stochastic}, we have:
\begin{equation}
    \begin{aligned}
        \nabla \log p(\Xs \mid \X_0) = -\frac{1}{\sqrt{s}\sigma_s} \mathbb{E}_{\X_0 \sim \pi(\X_0)} [\blz \mid \X_s],
    \end{aligned}
    \label{eq:stein}
\end{equation}
where $\X_s=\alpha_s \X_0 + \beta_s \X_1 + \sqrt{s} \sigma_s \blz$. In addition, 
\begin{equation}
    \begin{aligned}
        \bs (\Xs, \X_0) &= \dot{\alpha}_s \X_0 + \dot{\beta}_s \mathbb{E}_{\X_0 \sim \pi(\X_0)}[\X_1 \mid \Xs] + \sqrt{s}\dot{\sigma}_s \mathbb{E}_{\X_0 \sim \pi(\X_0)}[\blz \mid \Xs], \\
        \Xs &= \alpha_s \X_0 + \beta_s \mathbb{E}_{\X_0 \sim \pi(\X_0)}[\X_1 \mid \Xs] + \sqrt{s}\sigma_s \mathbb{E}_{\X_0 \sim \pi(\X_0)}[\blz \mid \Xs].
    \end{aligned}
    \label{eq:relate_drift_score_eqs}
\end{equation}
Solve $\mathbb{E}_{\X_0 \sim \pi(\X_0)}[\blz \mid \Xs]$ from \Cref{eq:relate_drift_score_eqs}: 
\begin{equation}
    \begin{aligned}
    \mathbb{E}_{\X_0 \sim \pi(\X_0)}[\blz \mid \Xs] = \frac{\beta_s \bs(\Xs, \X_0)-\dot{\beta}_s \Xs - (\beta_s \dot{\alpha}_s - \dot{\beta}_s \alpha_s)\X_0}{\sqrt{s}(\dot{\sigma}_s \beta_s - \sigma_s \dot{\beta}_s)}.
    \end{aligned}
    \label{eq:exp_of_z}
\end{equation}
Define $\As=\frac{1}{\sqrt{s}(\dot{\sigma}_s \beta_s - \sigma_s \dot{\beta}_s)}$ and $\cs(\Xs, \X_0)=\dot{\beta}_s \Xs + (\beta_s \dot{\alpha}_s - \dot{\beta}_s \alpha_s)\X_0$, and insert \Cref{eq:exp_of_z} into \Cref{eq:stein}, we relate the score function $\nabla \log \, p(\Xs \mid \X_0)$ and the drift $\bs(\Xs, \X_0)$ \cite{pmlr-v235-chen24n} by:
\begin{equation}
    \begin{aligned}
        \nabla \log \, p(\Xs, \X_0) = \As [\beta_s \bs(\Xs, \X_0) - \cs(\Xs, \X_0)].
    \end{aligned}
\end{equation}
\subsection{Estimation of the Gradient Log Likelihood of Observation}
\label{sec:estimate_loglikelihood}
The term \(\nabla \log p(\y \mid \Xs, \X_0)\) in \Cref{eq:bs_g_bayes} captures the observation information.
Since the observation model only directly links
$\y = \y_{k+1}$ and $\X_1 = \xnext$,
we compute this term by integrating with respect to $\X_1$: 
\begin{equation}
\begin{aligned}
  \nabla \log p(\y \mid \Xs, \X_0) &= \frac{\nabla \,p(\y|\Xs, \X_0)}{p(\y|\Xs, \X_0)} 
  =\frac{\nabla \int p(\y|\X_1)p(\X_1|\Xs, \X_0) \der \X_1 }{\int p(\y|\X_1)p(\X_1|\Xs, \X_0) \der \X_1 }\\
  &=\frac{\nabla \,\mathbb{E}_{\X_1  \sim p(\X_1 \mid \Xs, \X_0)}[p(\y|\X_1)]}{\mathbb{E}_{\X_1  \sim p(\X_1 \mid \Xs, \X_0)}[p(\y|\X_1)]}. 
\end{aligned}
\label{eq:jensen_estimation}
\end{equation}
In practice, we approximate the above expectations
by $J$ Monte Carlo samples,
$\X_1^{(j)} \sim p(\X_1 \mid \Xs, \X_0)$: 
\begin{equation}
    \begin{aligned}
        \mathbb{E}_{\X_1  \sim p(\X_1 \mid \Xs, \X_0)}[p(\y|\X_1)] \approx \frac{1}{J} \sum_{j=1}^{J} p(\y \mid \X_1^{(j)}).
    \end{aligned}
    \label{eq:monte_carlo}
\end{equation}
Plug \Cref{eq:monte_carlo} into \Cref{eq:jensen_estimation}:
\begin{equation}
\begin{aligned}
    \nabla \log p(\y \mid \Xs, \X_0) &\approx \frac{\nabla \,\frac{1}{J} \sum_{j=1}^{J} p(\y \mid \X_1^{(j)})}{\frac{1}{J} \sum_{j=1}^{J}  p(\y \mid \X_1^{(j)})}
  =\frac{\sum_{j=1}^{J}  p(\y \mid \X_1^{(j)}) \,\nabla \, \log p(\y \mid \X_1^{(j)})}{\sum_{j=1}^{J}  p(\y \mid \X_1^{(j)})}\\
  &=\sum_{j=1}^{J} w_j \nabla \, \log p(\y \mid \X_1^{(j)}),
\end{aligned}
\label{eq:jensen_estimation_mc_append}
\end{equation}
where we apply
a softmax function to the $J$ scalars
$\{\log p(\y \mid \X_1^{(j)})\}_{j=1}^{J}$
to compute the sample weights
$w_j = p(\y \mid \X_1^{(j)})/\sum_{j=1}^{J}  p(\y \mid \X_1^{(j)})$. 

\section{Implementation Details}
\label{sec:implementation}

\subsection{Algorithms}
\paragraph{Training}
We used a dataset consisting of multiple simulated state trajectories $\x_{0:K}$
to train the drift function
$\bs (\Xs, \X_0)$ in stochastic interpolants
(see \Cref{sec:training_details} for more details).
We approximate the cost function in \Cref{eq:objective_bs}
by the empirical loss: 
\begin{equation}
    \cL_b^{\text{emp}}(\hat{\blb}) = \frac{1}{K'} \sum_{k \in \cB_{K'}} \int_{0}^{1}
    \ell \bigl( \hat{\blb}_s(\Is^k, \xk), \blR_s^k \bigr) \, \der s,
    \label{eq: training objective}
\end{equation}
where $\ell(\cdot, \cdot)$ denotes a user-chosen discrepancy metric, depending on the experiment. $\cB_{K'} \subset \{0 : K\}$ is a subset of indices of cardinality $K^\prime \leq K$, with 
\begin{equation}
\begin{aligned}
    \Is^k &= \alpha_s \xk + \beta_s \xnext + \sqrt{s} \sigma_s \blz_k \\
    \blR_s^k &= \dot{\alpha}_s \xk + \dot{\beta}_s \xnext + \sqrt{s} \dot{\sigma}_s \blz_k,
\end{aligned}
\label{eq: training objective para}
\end{equation}
where $\blz_k \sim \cN(0, \I_D)$ and satisfy $\Ws \overset{d}{=} \sqrt{s} \blz$ with $\blz \sim \cN(0, \I_D)$ for all $s\in [0, 1]$.
We approximate the integral over $s$ in \Cref{eq: training objective}
via an empirical expectation sampling from $s \sim U([0,1])$. \Cref{alg:training} provides a detailed description of the model training process.

\paragraph{Inference} 
Our inference procedure generates trajectories conditioned on observations $\bm y_{L:K}$
using the learned drift model $\hat{\blb}_s(\Xs, \X_0)$.
We start by setting a specific state \xk as initial $\X_0$
and iterate over a predefined temporal grid $s_0 = 0 < s_1 < \cdots < s_N = 1$.
Within each iteration, we first compute posterior estimates
$\{\hat{\X}^{(j)}_{1}\}^{J}_{j=1}$
using \Cref{eq:1st_integration,eq:2nd_integration} for $J$ times.
Then, we move one step further towards $s_N$ on $\X_{s_n}$
by solving \Cref{eq:bs_g_bayes}
which involves backpropagating the gradient
$\nabla_{\X_{s_n}} \sum_{j=1}^{J} w_j \| \y - \cA(\hat{\X}^{(j)}_1) \|_2^2$
to enforce consistency with observations \yk.
We iterated this process autoregressively,
by setting $\X_{s_0}^{k+1}=\X_{s_N}^{k}$.
Empirically, we found that using a constant step size $\zeta_n$ across the inference process
produced generally good results, although fine-tuning $\zeta_n$ at each step
can slightly improve the performance \cite{siyichen2024blindinverse,chung2022diffusion}. The chosen $J$ values crossed all experiments, reported in \Cref{table:hyperparameters}, were sufficient large for stable performance, with subtle variation across the dimensionality of problems. An ablation study, detailed in \Cref{sub: Monte-carlo hyperparameters estimation}, 
further confirmed that larger $J$ offers diminishing returns.
Overall, 
\Cref{alg:sampling} summarizes the inference process.

\begin{algorithm}[t]
    \caption{Training} 
    \label{alg:training}
    \begin{algorithmic}[1]
       \STATE {\bfseries Input:} Dataset $\x_{0:K}$; minibatch size $K' \leq K$; coefficients $\alpha_s$, $\beta_s$, $\sigma_s$
       \REPEAT
       \STATE Compute $\Is^k$ and $\blR_s^k$ using \eqref{eq: training objective para} for $k \in \cB_{K'}$
       \STATE Compute the empirical loss $\cL_b^{\text{emp}}(\hat{\blb})$ in  \Cref{eq: training objective}
       \STATE Take the gradient step on $\cL_b^{\text{emp}}(\hat{\blb})$ to update $\hat{\blb}_s$
       \UNTIL{converged}
       \RETURN drifts $\hat{\blb}_s$
    \end{algorithmic}
\end{algorithm}

\begin{algorithm}[t]
    \caption{Inference} 
    \label{alg:sampling}
    
    \begin{algorithmic}[1]
    \STATE {\bfseries Input:} Observation $\y_{L:K}$, the measurement map \cA, initial state $\x_0$, model $\hat{\blb}_s(\X, \X_0)$, noise coefficient $\sigma_s$, grid $s_0 = 0 < s_1 < \cdots < s_N = 1$, i.i.d. $\blz_n \sim \cN(0, \I_D)$ for $n = 0:N-1$, step size $\zeta_n$, Monte Carlo sampling times $J$
    \STATE Set $\hat{\x}_{L-1} \leftarrow \x_{L-1}$
    \STATE Set the $(\Delta s)_n = s_{n+1} - s_{n}, \ n = 0 :N-1$
    \FOR{$k=L-1$ {\bfseries to} $K-1$} 
        \STATE $\X_{s_0}, \y \leftarrow \hat{\x}_{k}, \ynext$ 
        \FOR{$n=0$ {\bfseries to} $N-1$}
            
            \STATE $\X_{s_{n+1}}^{\prime} = \X_{s_n} + \hat{\blb}_s(\X_{s_n}, \X_{s_0})(\Delta s)_n + \sigma_{s_n}\sqrt{(\Delta s)_n}\blz_n$
            \STATE {$\{\hat{\X}^{(j)}_{1}\}^{J}_{j=1} \leftarrow$
            {Posterior estimation} ($\hat{\blb}_s$, $s_n$, $\X_0$, $\X_{s_n}$)} 
            \STATE {$\{w_j\}^{J}_{j=1} \leftarrow$
            Softmax$\left(\{\| \y - \cA(\hat{\X}^{(j)}_1) \|_2^2\}^{J}_{j=1}\right)$}
            \STATE {$\X_{s_{n+1}} = \X_{s_{n+1}}^{\prime} -\zeta_n \nabla_{\X_{s_n}} \sum_{j=1}^{J} w_j \| \y - \cA(\hat{\X}^{(j)}_1) \|_2^2$}
        \ENDFOR
        \STATE $\hat{\x}_{k+1} \leftarrow \X_{s_N}$
    \ENDFOR
    \RETURN \( \{\hat{\x}_k \}_{k=L}^K \)
    \end{algorithmic}
    
\end{algorithm}

\subsection{Constructing Training Dataset}
\label{sec:training_details}

\begin{figure}[t]
    \centering
    \includegraphics[width=0.7\linewidth]{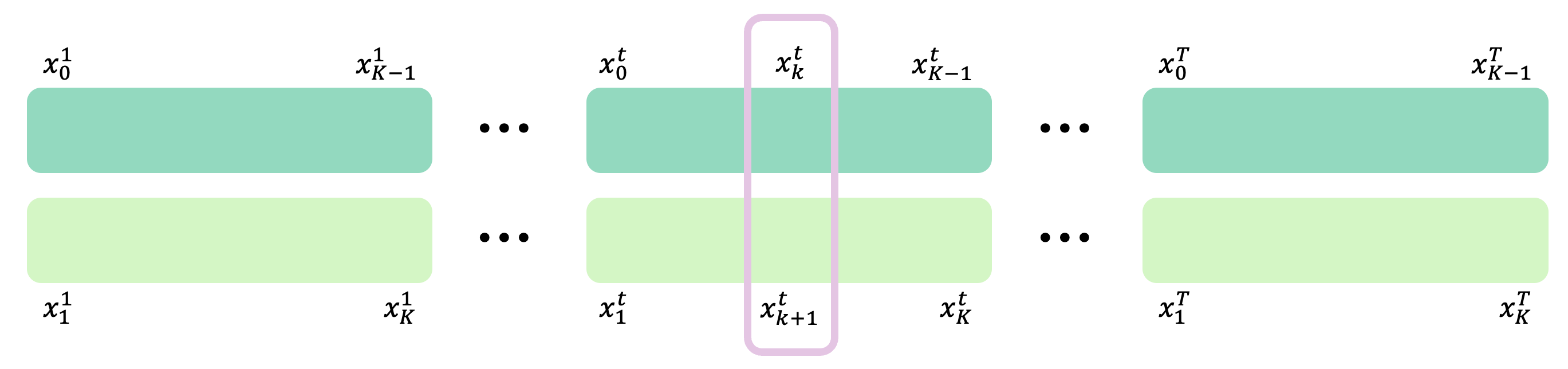}
    \vspace{-0.1in}
    \caption{\textbf{Structure of training data}. Consecutive states are paired across multiple simulated trajectories to construct the $\Is$ and $\Rs$ defined in \Cref{eq: training objective} for training the velocity model $\bs$.}
    \label{fig:training_data}
\end{figure}

In the training stage for all experiments, we require pairs of consecutive states to train the velocity model $\bs$. To generate these pairs, we proceed as follows:
\begin{enumerate}[leftmargin=*]
    \item \textbf{Trajectory simulation.} We simulate $T$ independent trajectories, denoted as $\x_{0:K}^{1:T}$, where each trajectory starts from a unique initial state $\x_0^t$. The state transitions within each trajectory follow the dynamics defined in \Cref{eq:stochastic_dynamics_model}.

    \item \textbf{Consecutive state pairs formation.} For each trajectory $t$, form two aligned sequences:
    \begin{enumerate}
        \item $\x_{0:K-1}^t$: The original sequence of states with each last state $x_{K}^t$ discarded.
        \item $\x_{1:K}^{t}$: The sequence of states shifted by one time step with each initial state $x_{0}^t$ discarded.
    \end{enumerate}
    The two sequences form pairs of consecutive states $(\xk^t, \xnext^t)$ for $k=0, 1, \cdots, K-1$. 

    \item \textbf{Concatenating trajectories.} Then, we concatenate $T$ sequences of $\x_{0:K-1}^t$ and $T$ sequences $\x_{1:K}^t$ end-to-end, respectively, into two long sequences: $\x_{0:K-1}^{1:T}$ and $\x_{1:K}^{1:T}$, where each state in the second sequence is the corresponding successor states in the first sequence. 

    \item \textbf{Sampling training data.} During training, batches of paired consecutive states are sampled. For each batch, we sample training data pairs as follows:
    \begin{enumerate}
        \item sample states from $\x_{0:K-1}^{1:T}$ as $\X_0$'s.
        \item retrieve the counterpart states from $\x_{1:K}^{1:T}$ as $\X_1$'s.
    \end{enumerate}
    The batches of state pairs $(\X_0, \X_1)$ are used to construct $\Is$ and $\Rs$ in \Cref{eq: training objective} for training the velocity model $\bs$ as described in \Cref{alg:training}.
\end{enumerate}
    
\noindent In summary, we draw many state pairs $(\X_0, \X_1)$ and estimate the integral over $s$ in \Cref{eq: training objective} by approximated via an empirical expectation over draws of $s \sim U(0,1)$, and for every $s$ and every  state pairs $(\X_0, \X_1)$, we independently compute \Cref{eq: training objective para} with $S$ samples of $\bm z_k$, so the training loss can be rewritten as:
    \begin{equation}
        \cL_b^{\text{emp}}(\hat{\blb}) = \frac{1}{K'} \sum_{k \in \cB_{K'}}\frac{1}{S} \sum_{s=1}^{S}
    \| \hat{\bm b}_s (\Is^k, \xk) - \Rs^k \|^2 \, ,
    \label{eq: training objective2}
    \end{equation}
where we randomly draws $s$ from $U(0,1)$. Note that the training batch size equals $S \times K'$. \Cref{fig:training_data} illustrates the structure of training data, showing how consecutive states are paired across trajectories.

\begin{table*}[t]
\begin{center}
\begin{sc}
\begin{tabular}{cclcc}
\hline
\multicolumn{2}{l}{}                &          & \shortstack{Monte Carlo\\Sampling Times $J$} & Sampling Step Size $\zeta$ \\ \hline
\multicolumn{3}{c}{Double-well}                & 17                         & 1         \\
\multicolumn{3}{c}{Lorenz 1963}                & 21                         & 0.0002      \\
\multirow{4}{*}{\begin{tabular}[c]{@{}c@{}}Incompressible \\ Navier Stokes \end{tabular}} & SR & 4x       & 25                         & 1         \\
                               & SR & 8x       & 25                         & 2         \\
                               & SO & 5\%      & 25                         & 1         \\
                               & SO & 1.5625\% & 25                         & 1.75      \\
\multicolumn{3}{c}{Particle Image Velocimetry}                        & 25                         & 1         \\ 
\multicolumn{3}{c}{Weather Forecasting} & 25 & 0.1 \\ 
\hline
\end{tabular}
\end{sc}
\end{center}
\caption{\textbf{Hyperparameters for FlowDAS in the inference stage} of all experiments  presented in this study. SR stands for super-resolution task and SO represents sparse observation (inpainting) task.}
\label{table:hyperparameters}
\end{table*}



\subsection{FlowDAS Inference Hyperparameters}
\label{sec:implementation_details}
We present key hyperparameters for FlowDAS in the inference stage, including the Monte Carlo sampling times $J$ and the sampling step size $\xi$ in \Cref{table:hyperparameters}. 



\section{Ablation Study}
\label{ablation_study}
This section examines different aspects of the proposed FlowDAS,
including an alternative to the method
and the hyperparameter settings.
We also provide various evaluation results.
To simplify notation,
we omit the explicit sampling distribution \( p(\x) \) in the expectation operator,
writing \( \mathbb{E}_{\x \sim p(\x)}[f(\x)] \)
simply as \( \mathbb{E}_{\x}[f(\x)] \).
The sampling distribution for \( \x \)
will be specified at the end of the equation where necessary.

\subsection{Monte Carlo Sampling and An Alternative}
In \Cref{sec:estimate_loglikelihood},
we estimate $\nabla \log p\,(\y \mid \Xs, \X_0)$ by
\begin{equation}
    \begin{aligned}
        \nabla \log p\,(\y \mid \Xs, \X_0)
        = \frac{\nabla \,\mathbb{E}_{\X_1}\left[p\,(\y|\X_1)\right]}
        {\mathbb{E}_{\X_1}\left[p\,(\y|\X_1)\right]},
    \end{aligned}
    \label{eq:likelihood_suppl}
\end{equation}
where we approximate the expectation term
by averaging $J$ Monte Carlo samples:
\begin{equation}
\begin{aligned}
    \mathbb{E}_{\X_1}\left[p\,(\y|\X_1)\right]
    \approx \frac{1}{J} \sum_{j=1}^{J} p\,(\y | \X_1^{(j)}), 
    \X_1^{(j)} \sim p\,(\X_1 | \Xs, \X_0).
\end{aligned}
\label{eq: MC bound analysis}
\end{equation}
This Monte Carlo estimate is an unbiased estimate
and the error is proportional to $\frac{1}{\sqrt{J}}$.
When the number of Monte Carlo samples, i.e., $J$,
is sufficiently large, the estimation error converges to zero.
Alternatively, one can also apply Jensen's inequality
to estimate $\nabla \log p\,(\y \mid \Xs, \X_0)$,
which provides 
a biased estimate:
\begin{equation}
\begin{aligned}
  \nabla \log p\,(\y \mid \Xs, \X_0) 
  &= \nabla \log \int p\,(\y|\X_1)p\,(\X_1|\Xs, \X_0) \der \X_1 \\
  &=\nabla \log \mathbb{E}_{\X_1} \left[p\,(\y|\X_1)\right] \geq \nabla \mathbb{E}_{\X_1} \left[\log p\,(\y|\X_1)\right].
\end{aligned}
\label{eq:jensen_estimation_suppl}
\end{equation}
The "$\geq$" arises from Jensen's inequality.  

\paragraph{Unbiased vs. biased estimation} 
\label{sub : The Jensen estimation used}
Let
$Z = p(\y|\X_1) = \frac{1}{\sqrt{2\pi}\gamma}e^{-\frac{(\y -\cA(\X_1))^2}{2\gamma^2}}$,
where $Z$ is a bounded random variable
within the finite range $[0,\frac{1}{\sqrt{2\pi}\gamma}]$.
As a result, the logarithm function is $\alpha\text{-Hölder}$ continuous
with $\alpha = 1$ and the gap introduced by Jensen's inequality,
i.e., the Jensen gap, can be explicitly bounded \cite{simic2008global} and given by
\begin{equation}
\begin{aligned}
    \left| \log \mathbb{E}_{\X_1}\left[p\,(\y|\X_1)\right] - \mathbb{E}_{\X_1}\left[\log p( \y|\X_1) \right]\right| \leq M\gamma^1_1
\end{aligned}
\label{eq: Jensen bound explicit}
\end{equation}
where $M$ is a constant that satisfying
$\left|\log Z - \log \mathbb{E}[Z] \right| \leq M \left|Z - \mathbb{E}[Z]\right|$
and $\gamma_1^1 = \mathbb{E} \left[| Z - \mathbb{E}\left[ Z \right] | \right]$.
Additionally, because $Z \in [0,\frac{1}{\sqrt{2\pi}\gamma}]$,
we have $M < 1$ and
$\gamma^1_1 \leq \max \left| Z-\mathbb{E}\left[Z\right] \right| \leq \frac{1}{\sqrt{2\pi}\gamma}$. 

\noindent And for this upper-bound $M\gamma^1_1$, when $s \rightarrow 1$, $p\,(\X_1 \mid \Xs,\X_0)$ will become a delta distribution concentrated on $\hat{\X}_1$ and $Z = p(\y|\X_1)$ will also have a delta distribution concentrated on $p(\y|\hat{\X}_1)$, and the $\gamma^1_1 \approx 0 $ finally. In conclusion, this bias should be controlled by 
\begin{equation}
\begin{aligned}
    \left| \log \mathbb{E}_{\X_1} \left[p\,(\y| \X_1)\right]  -    \mathbb{E}_{\X_1} \left[\log p\,(\y|\X_1)\right] \right| \leq \frac{1}{\sqrt{2\pi}\gamma}
\end{aligned}
\label{eq: Jensen bound explicit2}
\end{equation}
Although this bias is theoretically bounded, it still results in a slight degradation in performance.
\Cref{unbiasvsbias}
shows the comparison for the Lorenz experiment
to illustrate this point.

\begin{table}[t]
\begin{center}
\begin{sc}
\begin{tabular}{clcc}
\toprule
& & unbiased& biased  \\
\midrule
$\uparrow$ &
$ \log p(\hat{\x}_{2:K} \mid \hat{\x}_1)$  & \textbf{17.29} & -36.98 \\
$\uparrow$ &
$\log p(\y \mid \hat{\x}_{1:K})$ & \textbf{-0.228} & -1.530 \\
$\downarrow$ &
$W_1(\x_{1:K}, \hat{\x}_{1:K})$   & \textbf{0.106} & 0.111\\
$\downarrow$ &
RMSE$(\x_{1:K}, \hat{\x}_{1:K})$  &  \textbf{0.202}      &  0.363       \\
\bottomrule
\end{tabular}
\caption{\textbf{Evaluation of unbiased vs. biased estimation}. 
Comparison of metrics between unbiased and biased estimation in the Lorenz experiments. The results demonstrate that the unbiased estimation outperforms the biased estimation.}
\label{unbiasvsbias}
\end{sc}
\end{center}
\end{table}

\paragraph{Hyperparameters for Monte Carlo Sampling}
\label{sub: Monte-carlo hyperparameters estimation}
We examine the estimation process and associated hyperparameters in \Cref{eq: MC bound analysis}, where the expectation is computed using a Monte Carlo method. The hyperparameter \( J \), is referred to as the number of Monte Carlo sampling iterations in \Cref{alg:sampling}. It is important to note that increasing \( J \) does not lead to an increase in neural network evaluations but only involves additional Gaussian noise simulations, which are computationally lightweight. To illustrate the effect of \( J \), we use numerical results from the Lorenz experiment. As shown in \Cref{n's eva}, increasing \( J \) can improve performance by approximately 20\%, with negligible impact on computational time.

\begin{table}[!ht]

\begin{center}
\begin{sc}
\begin{tabular}{lcccccc}
\toprule
$J=$ & 3 & 6&12 & 21 & 30 & 50 \\
\midrule
RMSE & 0.167 & 0.148 & 0.153 & 0.142 & 0.150 & 0.138 \\
\bottomrule
\end{tabular}
\end{sc}
\end{center}
\caption{\textbf{Effect of $J$}. The RMSE of generated state trajectories for FlowDAS is evaluated with different Monte Carlo sampling times $J$ in \Cref{eq: MC bound analysis} for the Lorenz 1963 experiment. As $J$ increases, the RMSE initially decreases, indicating improved performance, and then stabilizes.} 
\label{n's eva}
\end{table}
\subsection{Posterior Estimation Methods}
\label{sub : The Method For Posterior Estimation}

We also evaluate the impact of different methods for posterior estimation: as defined in \Cref{eq:1st_integration,eq:2nd_integration}. The results are presented in \Cref{posterior method's eva}, where ``1st-order'' and ``2nd-order'' refer to 1st-order Milstein method and 2nd-order stochastic Runge-Kutta method, respectively. ``No correction'' indicates forecasting purely based on the model without incorporating observation information. The results show that both 1st-order and 2nd-order estimations provide reasonable accuracy. However, the 2nd-order estimation consistently delivers better performance. This suggests that employing more accurate estimations of \( p\,(\X_1 | \Xs, \X_0) \) can effectively enhance model performance. Beyond the 2nd-order method, higher-order approaches like the Runge-Kutta 4th-order (RK4) method could further improve accuracy. However, these methods come with increased computational cost: 2nd-order estimation requires two neural network evaluations per step compared to one for 1st-order estimation, while RK4 requires four evaluations per step. In our experiments, we find that the 2nd-order estimation strikes a good balance between performance and efficiency, making it a practical choice. Further exploration of higher-order methods will be left for future research.

\begin{table}[t]
\begin{center}
\begin{sc}
\begin{tabular}{lcccccc}
\toprule
     & 1st-order & 2nd-order & No correction  \\
\midrule
$32^2 \rightarrow 128^2$ & 0.048 & \textbf{0.038 }& 0.206 \\
$16^2 \rightarrow 128^2 $ & 0.101 & \textbf{0.067} & 0.206 \\
\bottomrule
\end{tabular}
\end{sc}
\end{center}
\caption{\textbf{Effect of posterior estimation}. The RMSE of vorticity estimate from FlowDAS is evaluated on the incompressible Navier-Stokes task using different posterior estimation methods as defined in \Cref{eq:1st_integration,eq:2nd_integration}, and forecasting without observations (i.e., no correction). For the super-resolution tasks $16^2 \rightarrow 128^2$ and $32^2 \rightarrow 128^2$, both posterior estimation methods significantly outperform forecasting without observations.  Among them, the 2nd-order method achieves the lowest RMSE.} 
\label{posterior method's eva}
\end{table}

\section{Additional Experiment Results}
\label{sec:experiments_details}
This section provides the details of our experiments,
including additional experiments and results.



\subsection{Double-well Potential Problem}
\label{sec:double_well}

\subsubsection{The Double-well Potential System}
\begin{figure}[ht]
\centering
\begin{subfigure}[t]{0.48\textwidth}
    \centering
    \includegraphics[width=\textwidth]{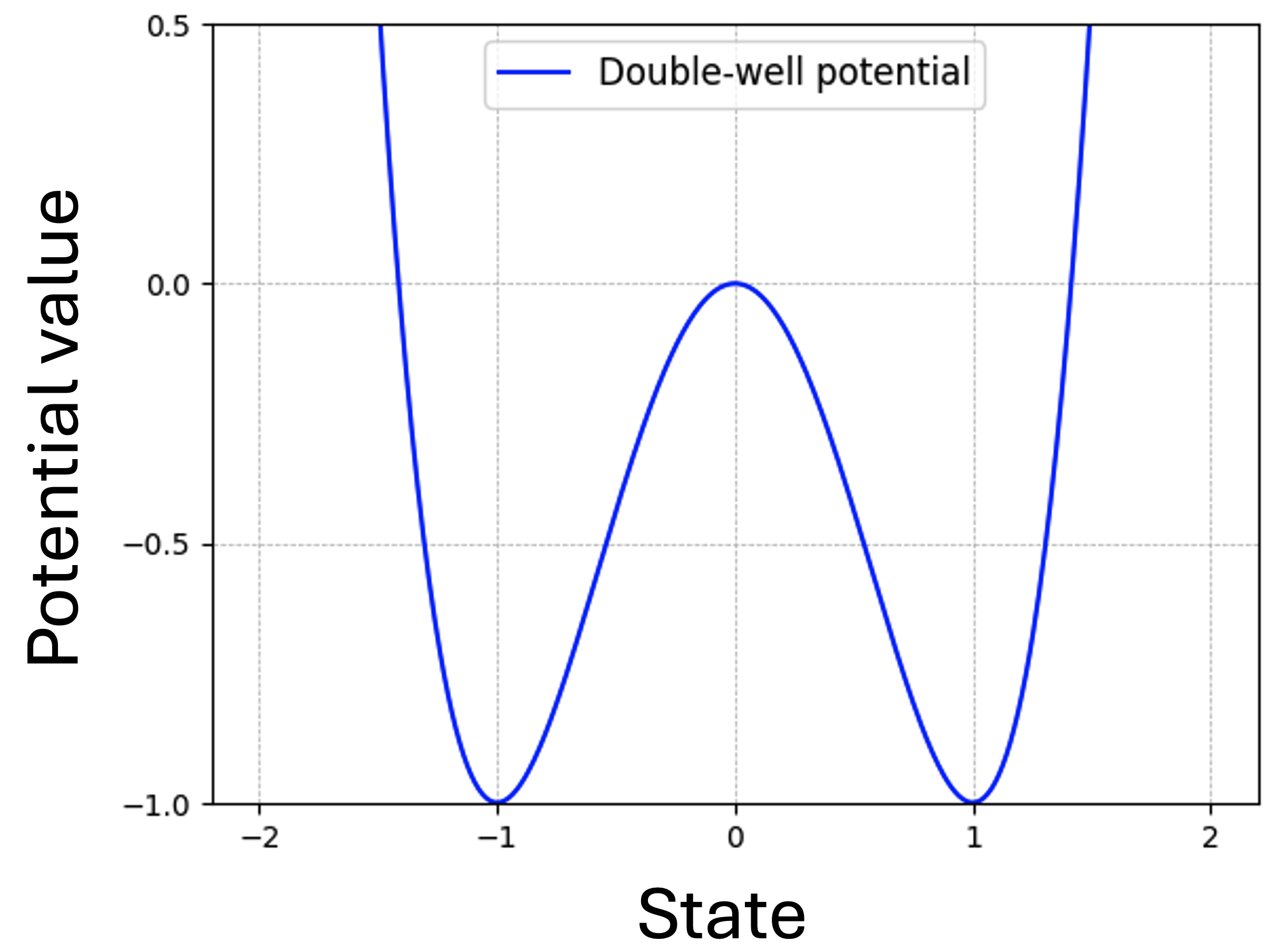}
    \caption{\textbf{Illustration of the Double-Well Potential Problem}. 
    In the double-well potential system, particles are typically captured at the bottoms of the wells (\( x = 1 \) and \( x = -1 \)). With low probability, particles can transition between wells due to the stochastic term in the transition equation. }
    \label{fig: double well potential illustration}
\end{subfigure}
\hfill
\begin{subfigure}[t]{0.48\textwidth}
    \centering
    \includegraphics[width=\textwidth]{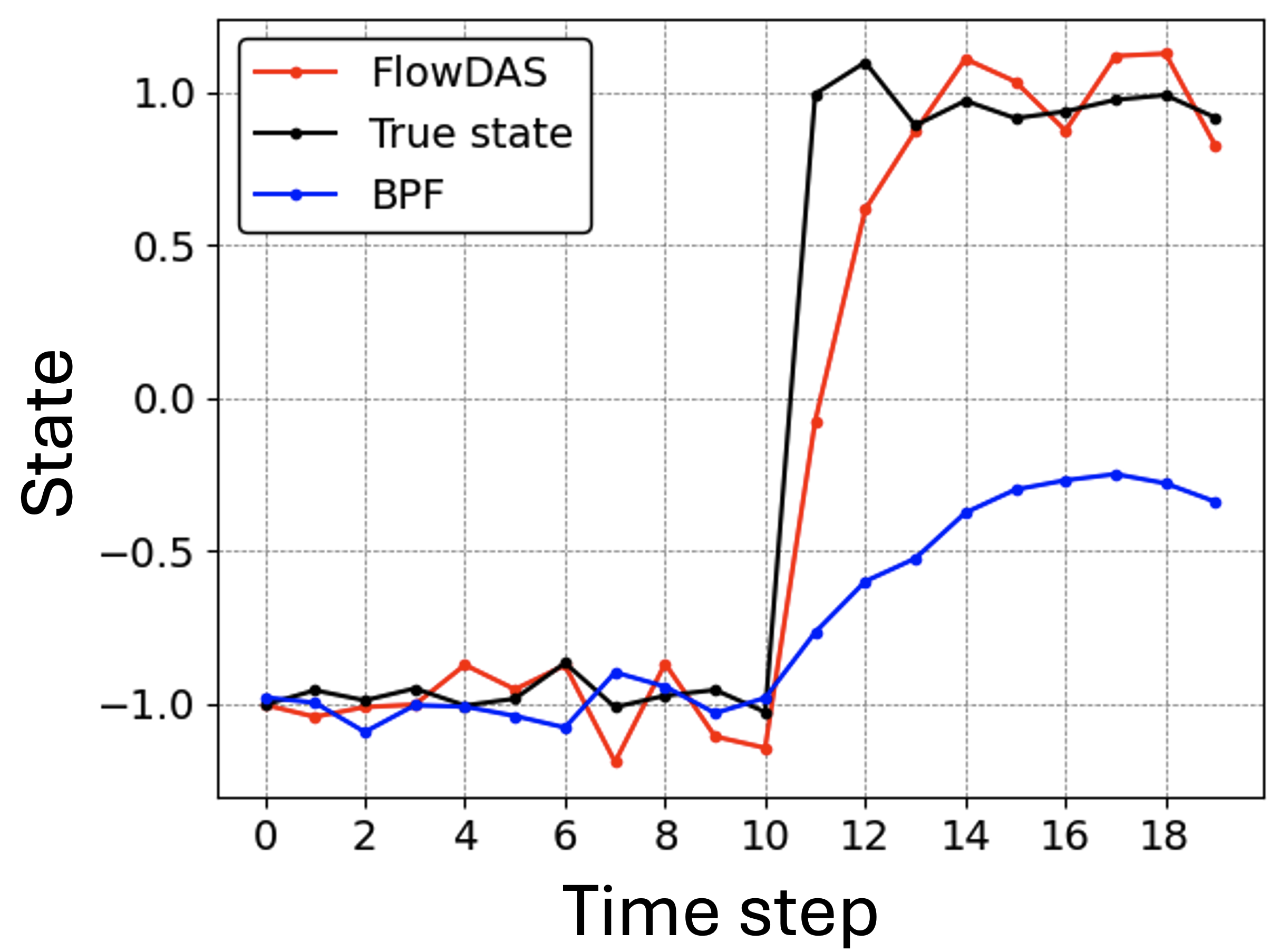}
    \caption{\textbf{Visualization Results of Double-Well Potential Problem}. We show the visualization results of FlowDAS and BPF, while the score-based solver SDA fails to produce reasonable results. FlowDAS can track the dramatic change of the particles that BPF struggles to immediately react to.}
    \label{double well visual results}
\end{subfigure}
\caption{Illustrations and visualization results for the double-well potential problem.}
\label{double well combined}
\end{figure}

In this experiment, we investigate a 1D tracking problem in a dynamical system driven by the double-well potential. The system is governed by the following stochastic dynamics:

\begin{equation}
   \der x = -4 x ( x^2 - 1) \der t + \beta_d \der \xi_t
\label{eq:double-well SDE}
\end{equation}
with observation model defined by
$\cA(x)=x^3$. The observations are given by
\begin{equation}
   y = \cA(x) + \eta =x^3 +\eta
\label{eq:double-well observation}
\end{equation}
where the stochastic force $\xi_t$ is a standard Brownian motion with diffusion coefficient of $\beta_d=0.2$. $\eta$ is standard Gaussian noise with standard deviation of $0.2$. The $\Psi$ is the derivative of the potential $U(x)= x^4-2 x^2+  f$, where $f$ is a function independent of $x$. The system describes a particle trapped in the wells located at $ x = 1$ and $ x=-1$, with small fluctuations around these points, as illustrated in \Cref{fig: double well potential illustration}.

\subsubsection{Training and Testing Data Generation}
We trained the model on simulated trajectories generated by numerically solving the transition equation using a temporal step size of $0.1$. The training dataset consisted of $500$ trajectories, each of length 100, with initial points uniformly sampled from the range $[-2, 2]$. In the testing stage, we introduced stronger turbulence to the system, causing the particle to occasionally switch wells ($x \rightarrow -x$).

\subsubsection{Neural Network for Learning the Drift Velocity}
In this low-dimensional double-well potential task, the drift velocity $\bs$ is approximated using a fully connected neural network with 3 hidden layers, each having a hidden dimension of 50. Both the input and output dimensions of the network are 1. For the condition $\X_0$ and timestep $s$, we empirically find that embedding $\X_0$, similar to how $s$ is embedded, outperforms directly using $\X_0$ as an additional input to the network. The intuition behind this approach is that embedding $\X_0$ helps the network better distinguish between the two variables, $\Xs$ and $\X_0$. The model is trained using the Adam optimizer with a base learning rate of $0.005$, along with a linear rate scheduler. Training is conducted for $5000$ epochs.

\subsubsection{Hyperparameters During Inference}
Hyperparameters of the inference procedure, specified for the double-well potential task, are presented in \Cref{table:hyperparameters}.

\subsubsection{Baseline}
\label{sec:double-well}
For this task, we compare our method with SDA and the classic BPF method. For SDA, we fix the window size to two and use the same training data as FlowDAS to ensure fairness. The local score network is implemented as a fully connected neural network, following the architecture proposed in \cite{bao2023scorebasednonlinearfilterdata}. The score network consists of 3 hidden layers, each with a hidden dimension of 50. The model is trained using the AdamW optimizer with a base learning rate of $0.001$, a weight decay of $0.001$, and a linear learning rate scheduler. Training is conducted for $5000$ epochs. For the BPF method, the particle density is set to $16384$.

\subsubsection{Results}
The visual results are shown in \Cref{double well visual results}. Surprisingly, while both FlowDAS and the classic BPF method produce reasonable results, FlowDAS demonstrates superior performance in tracking dramatic changes in the trajectory. In contrast, the score-based solver SDA fails to produce reasonable results. This failure arises because the starting point of SDA during optimization is purely Gaussian noise, leading to a poor initial estimation of the target state. Furthermore, the cubic observation model amplifies the differences, causing the optimization gradients to explode and resulting in recovery failure.

In FlowDAS, the error is bounded by \( ||\y_{k+1} - \cA(\xk)||_2^2 \) because the generation process begins with the previous state, which serves as a proximal estimate of the target state. Consequently, FlowDAS undergoes a more stable optimization process.

Compared to the classic BPF method, we find that it struggles to capture dramatic changes in the trajectory. This limitation is primarily due to the small diffusion coefficient \( \beta_d \), such as \( \beta_d = 0.2 \), which causes the predicted filtering density in BPF to concentrate around the mean value dictated by the deterministic part of the transition equation. As a result, extreme cases lying in the tail of the future state distribution \( p(\xnext \mid \xk) \) are often missed. This phenomenon can be explained by the truncation error arising from the finite particle space \cite{bao2023scorebasednonlinearfilterdata}.

In contrast, FlowDAS is capable of immediately sampling from the true conditional distribution as the steps become sufficiently large. This allows FlowDAS to better capture the tail region of the true distribution and react to dramatic changes, even those in low-probability areas. Additionally, FlowDAS can effectively incorporate observation information, enabling it to balance prediction and observation even when the true underlying state occurs in a low-probability region.


\subsection{Lorenz 1963}
\label{sec:supp in lorenz}
\subsubsection{The Lorenz 1963 Dynamic System}
\label{subsub: RK4 lorenz}
To simulate this system, we use the RK4 method, which updates the solution from time \( t_n \) to \( t_{n+1} = t_n + h \) using the following formulas for each variable \( a \), \( b \), and \( c \):
\begin{equation}
\begin{aligned}
    k_{1a_{n}} &= h \cdot \mu (b_{n} - a_{n}), \\
    k_{1b_{n}} &= h \cdot \left( a_{n} (\rho - c_{n}) - b_{n} \right), \\
    k_{1c_{n}} &= h \cdot \left( a_{n} b_{n} - \tau c_{n} \right),
\end{aligned}
\label{eq: rk1}
\end{equation}

\begin{equation}
\begin{aligned}
    k_{2a_{n}} &= h \cdot \mu \left( (b_{n} + \frac{k_{1b_{n}}}{2}) - (a_{n} + \frac{k_{1a_{n}}}{2}) \right), \\
    k_{2b_{n}} &= h \cdot \left( (a_{n} + \frac{k_{1a_{n}}}{2})(\rho - (c_{n} + \frac{k_{1c_{n}}}{2})) - (b_{n} + \frac{k_{1b_{n}}}{2}) \right), \\
    k_{2c_{n}} &= h \cdot \left( (a_{n} + \frac{k_{1a_{n}}}{2})(b_{n} + \frac{k_{1b_{n}}}{2}) - \tau (c_{n} + \frac{k_{1c_{n}}}{2}) \right),
\end{aligned}
\label{eq: rk2}
\end{equation}
\begin{equation}
\begin{aligned}
    k_{3a_{n}} &= h \cdot \mu \left( (b_{n} + \frac{k_{2b_{n}}}{2}) - (a_{n} + \frac{k_{2a_{n}}}{2}) \right), \\
    k_{3b_{n}} &= h \cdot \left( (a_{n} + \frac{k_{2a_{n}}}{2})(\rho - (c_{n} + \frac{k_{2c_{n}}}{2})) - (b_{n} + \frac{k_{2b_{n}}}{2}) \right), \\
    k_{3c_{n}} &= h \cdot \left( (a_{n} + \frac{k_{2a_{n}}}{2})(b_{n} + \frac{k_{2b_{n}}}{2}) - \tau (c_{n} + \frac{k_{2c_{n}}}{2}) \right),
\end{aligned}
\label{eq: rk3}
\end{equation}
\begin{equation}
\begin{aligned}
    k_{4a_{n}} &= h \cdot \mu \left( (b_{n} + k_{3b_{n}}) - (a_{n} + k_{3a_{n}}) \right), \\
    k_{4b_{n}} &= h \cdot \left( (a_{n} + k_{3a_{n}})(\rho - (c_{n} + k_{3c_{n}})) - (b_{n} + k_{3b_{n}}) \right), \\
    k_{4c_{n}} &= h \cdot \left( (a_{n} + k_{3a_{n}})(b_{n} + k_{3b_{n}}) - \tau (c_{n} + k_{3c_{n}}) \right).
\end{aligned}
\label{eq:rk4}
\end{equation}
Then, the updates for $a_{n}$, $b_{n}$, and $c_{n}$ are:
\begin{equation}
\begin{aligned}
    a_{n+1} &= a_{n} + \frac{1}{6}(k_{1a_{n}} + 2k_{2a_{n}} + 2k_{3a_{n}} + k_{4a_{n}}), \\
    b_{n+1} &= b_{n} + \frac{1}{6}(k_{1b_{n}} + 2k_{2b_{n}} + 2k_{3b_{n}} + k_{4b_{n}}), \\
    c_{n+1} &= c_{n} + \frac{1}{6}(k_{1c_{n}} + 2k_{2c_{n}} + 2k_{3c_{n}} + k_{4c_{n}}).
\end{aligned}
\end{equation}
After solving these deterministic updates, the stochastic force $(\xi_1, \xi_2, \xi_3)$ is added to $(a_{n+1}, b_{n+1}, c_{n+1})$, which leads to the Lorenz 1963 system dynamic equations described in \Cref{eq:lorenz transition}. 

\begin{figure*}[t]
    \centering
    \includegraphics[width=\textwidth]{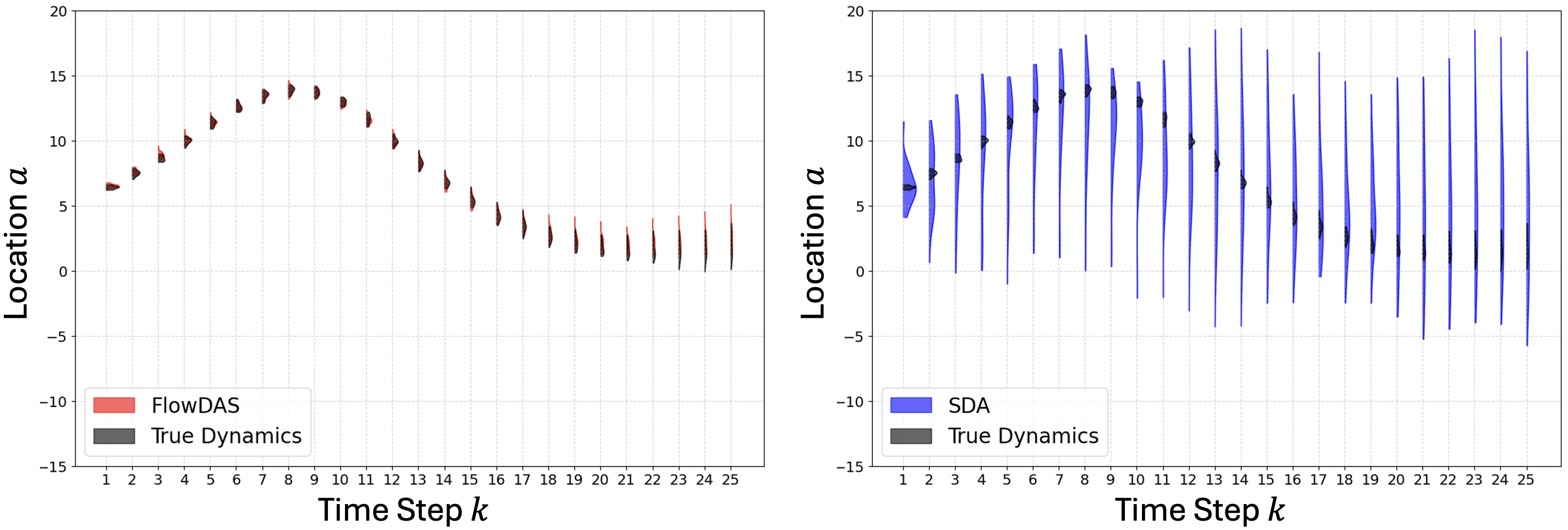}
    \caption{\textbf{Forecaseting dynamics modeling evaluation in the Lorenz 1963 task: FlowDAS vs. SDA}. FlowDAS produces results that closely match the true states, demonstrating its ability to learn the underlying transition dynamics \( p(\xnext \mid \xk) \). In contrast, SDA exhibits rapid divergence from the true states. This divergence arises because SDA focus on modeling the joint distribution \( p(\xnext, \xk) \) rather than directly learn the transition dynamics \( p(\xnext \mid \xk) \), which is inherently less suitable for capturing the system's underlying dynamics.}
    \label{fig:dynamics_modeling}
\end{figure*}

\subsubsection{Training and Testing Data Generation}
We apply the RK4 method described in \Cref{subsub: RK4 lorenz} to generate the simulated training and testing data.

\subsubsection{Hyperparameters During Inference}
Hyperparameters of the inference procedure, specified for the Lorenz 1963 task, are presented in \Cref{table:hyperparameters}. Noticeably, the step size $\zeta$ is smaller than in other experiments due to the chaotic nature of the Lorenz 1963 system, which requires finer step size to accurately capture its dynamics.

\subsubsection{Neural Network for Learning the Drift Velocity}
In this task, we use a fully connected neural network with 5 hidden layers, each with a hidden dimension of 256, to approximate the velocity field \( \bs\). The input and output dimensions are both 3. For the condition \( \X_0 \) and timestep \( s \), we use embeddings of dimension 4. The model is trained using the Adam optimizer with a base learning rate of \( 0.005 \), and a linear learning rate scheduler is applied. Training is conducted over \( 23000 \) epochs.

\subsubsection{Baseline}
We compare FlowDAS with SDA and BPF. For the SDA, we use a score neural network with a fixed window size of $2$. This local score function is implemented using a fully connected neural network with 5 hidden layers, each having a hidden dimension of $256$. The model is trained using the AdamW optimizer with a base learning rate of $0.001$ and a linear scheduler over $23000$ epochs. For BPF, the particle density is set to $16384$.

\subsubsection{Accuracy of the Learned Dynamics}
We evaluate the dynamic learning performance of FlowDAS and SDA, focusing on their ability to forecast future states \textbf{without} using observations. For BPF, we do not evaluate its performance in this context since it explicitly incorporates the true system dynamics. 

As SDA is designed to rely on observations and does not work in an auto-regressive manner, we adapt it for this evaluation by using the previous state as a pseudo-observation. The SDA then forecasts the next state based on this input. For FlowDAS, we disable the observation-based update step by setting the step size \( \zeta_n = 0 \).

Both methods are initialized with the same initial state, and we simulate 64 independent trajectories of length 25. True states are generated by solving \Cref{eq:lorenz transition} using the RK4 method, starting from the same initial point. The visualization of location \( a \) across 25 time steps is shown in \Cref{fig:dynamics_modeling}.


\subsection{Incompressible Navier-Stokes Flow}
\label{sec:ns_rest}

\subsubsection{Incompressible Navier-Stokes Flow Problem Settings}
\label{subsub: problem settings of ns}
For the incompressible Navier-Stokes flow problem, we adopted the problem setting from \cite{pmlr-v235-chen24n} and used the provided training data. The choices of experiment parameters are as follows: \( \nu = 10^{-3} \), \( \alpha = 0.1 \), \( \varepsilon = 1 \). The stochastic force \( \xi \) is defined as:
\begin{equation}
\begin{aligned}
    \xi(t, x, y) & = W_1(t) \sin(6x) + W_2(t) \cos(7x) + W_3(t) \sin(5(x + y)) + W_4(t) \cos(8(x + y)) \\
    & + W_5(t) \cos(6x) + W_6(t) \sin(7x) + W_7(t) \cos(5(x + y)) + W_8(t) \sin(8(x + y))
\end{aligned}
\label{eq:rk2}
\end{equation}
For more details and insights about this problem, see \cite{pmlr-v235-chen24n}.

\subsubsection{Hyperparameters During Inference}
\Cref{table:hyperparameters} presents the
hyperparameters of the inference procedure
for the incompressible Navier-Stokes flow task.

\subsubsection{Neural Network for Learning the Drift Velocity}
\label{subsub sec:details about the ns neural network}

We use a U-Net architecture to approximate \( \bs \), following the network proposed in \cite{pmlr-v235-chen24n}. The conditioning on \( \X_0 \) is implemented through channel concatenation in the input. The architectural details are as follows:
\begin{itemize}[leftmargin=*]
    \item {Number of states conditioned on}: 10.
    \item {Number of initial channels}: 128.
    \item {Multiplication factor for the number of channels at each stage}: (1, 2, 2, 2).
    \item {Number of groups for group normalization in ResNet blocks}: 8.
    \item {Dimensionality of learned sinusoidal embeddings}: 32.
    \item {Dimensionality of each attention head in the self-attention mechanism}: 64.
    \item {Number of attention heads in the self-attention layers}: 4.
\end{itemize}

We employ the AdamW optimizer \cite{Loshchilov2017DecoupledWD} with a cosine annealing schedule to reduce the learning rate during training. The base learning rate is set to \( 2 \times 10^{-4} \). Training is conducted with a batch size of $32$ for $1000$ epochs.

\subsubsection{Baseline}
We compare FlowDAS to FNO-DA, Transolver-DA, and SDA.

\paragraph{FNO-DA}
The model is configured with $7$ stacked Fourier layers. Each layer processes data in a $128$-dimensional feature space and truncates the Fourier series to include $6$ modes in each spatial direction. A padding of $6$ is applied before the Fourier operations and removed afterward. The model is provided with the previous $10$ states to predict the state at the next time step. We train the model for $1000$ epochs, using a batch size of $64$ and the AdamW optimizer with an initial learning rate of $0.0001$.

\paragraph{Transolver-DA} 
The model has $8$ layers of transolver blocks whose hidden channels are $128$ and have $32$ slices, with $8$ attention heads. The model is provided with the previous $10$ states to predict the state at the next time step. We train the model for $1000$ epochs using a batch size of $64$, the AdamW optimizer, and a OneCycleLR learning rate scheduler with an initial learning rate of  $5e^{-4}$.

\paragraph{SDA}
The score neural network is configured with a temporal window of $11$, of which the first $10$ inputs are conditions and the forecast horizon is $1$; an embedding layer dimensionality of $32$; and a base number of feature channels $256$.
The network depth is $5$, and the activation function used is SiLU \cite{Loshchilov2017DecoupledWD}. 
We train the model using a batch size of $64$ and the AdamW optimizer, with a linear learning rate scheduler (initial learning rate $0.001$) and a weight decay of $0.001$.

During inference, the $\alpha_1$ and $\alpha_2$ used is varied from task to task for both the FNO-DA and the Transolver-DA. We show the detailed parameter settings in \Cref{table:hyperparameters 4 transolver,table:hyperparameters 4 FNO}. We use an AdamW optimizer for Kalman update with a learning rate of $1e^{-4}$ and a maximum iteration time of $2000$ to ensure the loss converges. Without other indications, we adopt the same optimizer settings for both FNO-DA and Transolver-DA.

\begin{table*}[t]
\begin{center}
\begin{sc}
\begin{tabular}{cclcc}
\hline
\multicolumn{2}{l}{}                &          &  $\alpha_1$ &  $\alpha_2$\\ \hline
\multirow{4}{*}{\begin{tabular}[c]{@{}c@{}}Incompressible \\ Navier Stokes \end{tabular}} & SR & 4x       & 0.63                         & 1       \\
                               & SR & 8x       & 1.59                         & 1         \\
                               & SO & 5\%      & 0.10                         & 1         \\
                               & SO & 1.5625\% & 0.29                         & 1      \\
\multicolumn{3}{c}{Weather Forecasting} & 1.22 & 1 \\ 
\hline
\end{tabular}
\end{sc}
\end{center}
\caption{\textbf{Hyperparameters for FNO-DA in the inference stage} of all experiments  presented in this study. SR stands for super-resolution task and SO represents sparse observation (inpainting) task}
\label{table:hyperparameters 4 FNO}
\end{table*}

\begin{table*}[t]
\begin{center}
\begin{sc}
\begin{tabular}{cclcc}
\hline
\multicolumn{2}{l}{}                &          &  $\alpha_1$ &  $\alpha_2$\\ \hline
\multirow{4}{*}{\begin{tabular}[c]{@{}c@{}}Incompressible \\ Navier Stokes \end{tabular}} & SR & 4x       & 0.53                         & 1       \\
                               & SR & 8x       & 1.21                         & 1         \\
                               & SO & 5\%      & 0.14                         & 1         \\
                               & SO & 1.5625\% & 0.22                         & 1      \\
\multicolumn{3}{c}{Weather Forecasting} & 5.00 & 1 \\ 
\hline
\end{tabular}
\end{sc}
\end{center}
\caption{\textbf{Hyperparameters for Transolver-DA in the inference stage} of all experiments  presented in this study. SR stands for super-resolution task and SO represents sparse observation (inpainting) task}
\label{table:hyperparameters 4 transolver}
\end{table*}

\subsubsection{Kinetic Energy Spectrum Analysis}
\label{subsubsec: detail of kinetic energy spectrum}

We evaluate the methods from a physics perspective using the kinetic energy spectrum. A better method produces results that align more closely with the kinetic energy spectrum of the true state. The kinetic energy spectrum is computed as follows: 
\begin{equation}
    E(k_n) = \sum_{\substack{p, q \\ k_{p, q} \, \in \, \text{bin } n}} E(k_{x_p}, k_{y_q}),
\end{equation}
where \( k_{p, q} \) represents the wavenumbers grouped into bin \( n \), \( E(k_{x_p}, k_{y_q}) \) is the kinetic energy at each wavenumber and $p$,$q$ are the index representing a specific discrete wavenumber along the $k_x$ or $k_y$ axis.

Since the direct outputs of both FlowDAS and SDA are vorticity fields, it is necessary to first convert the vorticity into velocity before computing the kinetic energy spectrum. This conversion is achieved by solving the Poisson equation \( \Delta \psi = -\omega \) to obtain \( \psi \), and then calculating its gradient \( \bm{v} = -\nabla \psi \) to derive the velocity \( \bm{v} \).

We present the results of the kinetic energy spectrum analysis for the Navier-Stokes flow super-resolution and sparse observation tasks.

\begin{figure*}[t]
    \centering
    \includegraphics[width=0.75\textwidth]{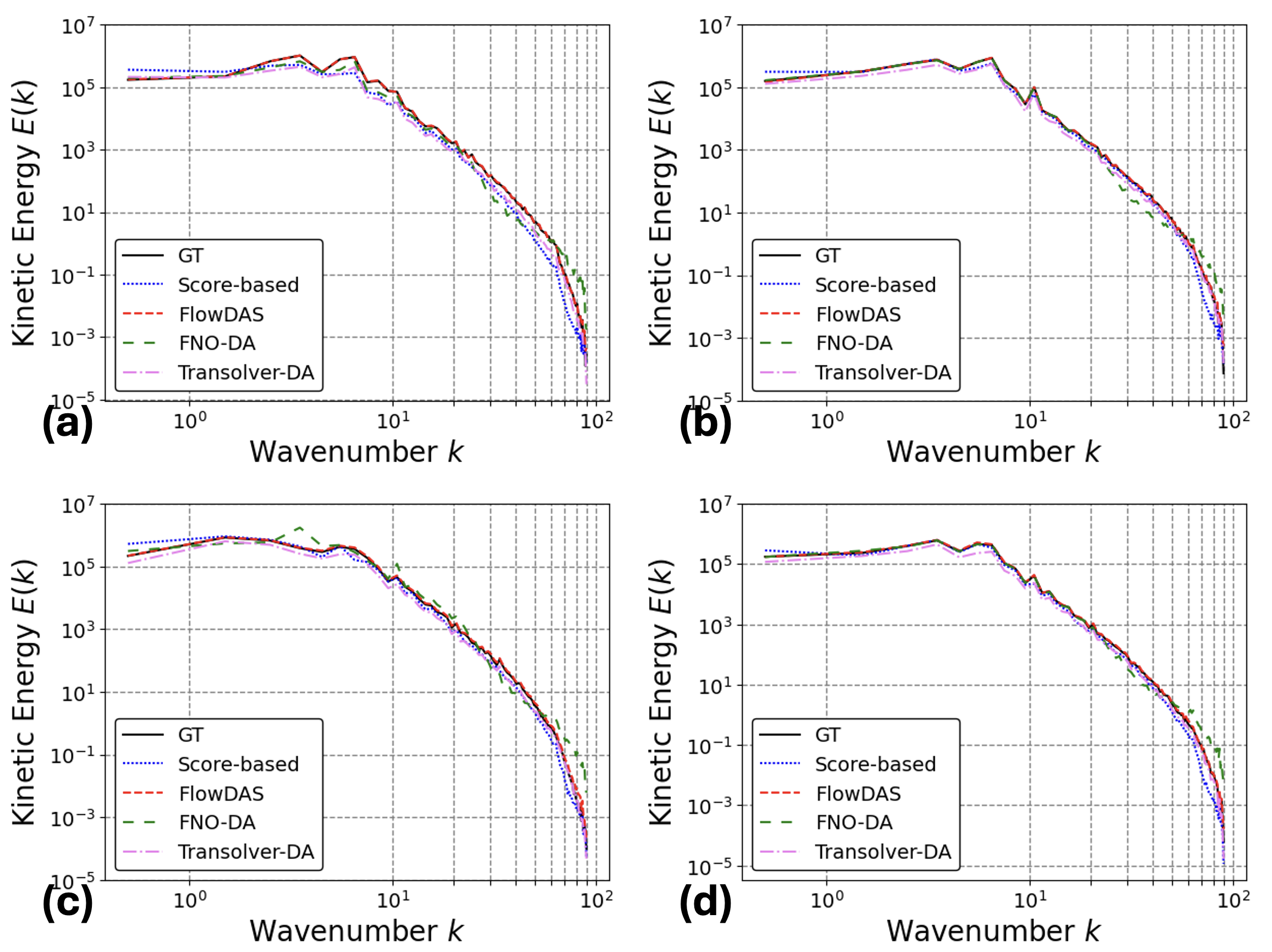}
    \caption{\textbf{The kinetic energy spectrum of:} (a) super-resolution task: $16^2 \rightarrow 128^2$; (b) super-resolution task: $32^2 \rightarrow 128^2$; (c) sparse observation task: 1.5625\%; (d) sparse observation task: 5\%, in the incompressible Navier-Stokes flow task. We present the kinetic energy spectrum of the true state alongside the estimations from FlowDAS and baseline methods. FlowDAS can produce results that better aligned with the true state in terms of the kinetic energy spectrum, evidenced by the oscillations in the spectrum of baselines, indicating FlowDAS's superiority in recovering the physics information and effectiveness as a surrogate model for stochastic dynamic systems.}
    \label{fig:spectral_4}
\end{figure*}

\begin{figure*}[ht!]
    \centering
    \includegraphics[width=0.75\textwidth]{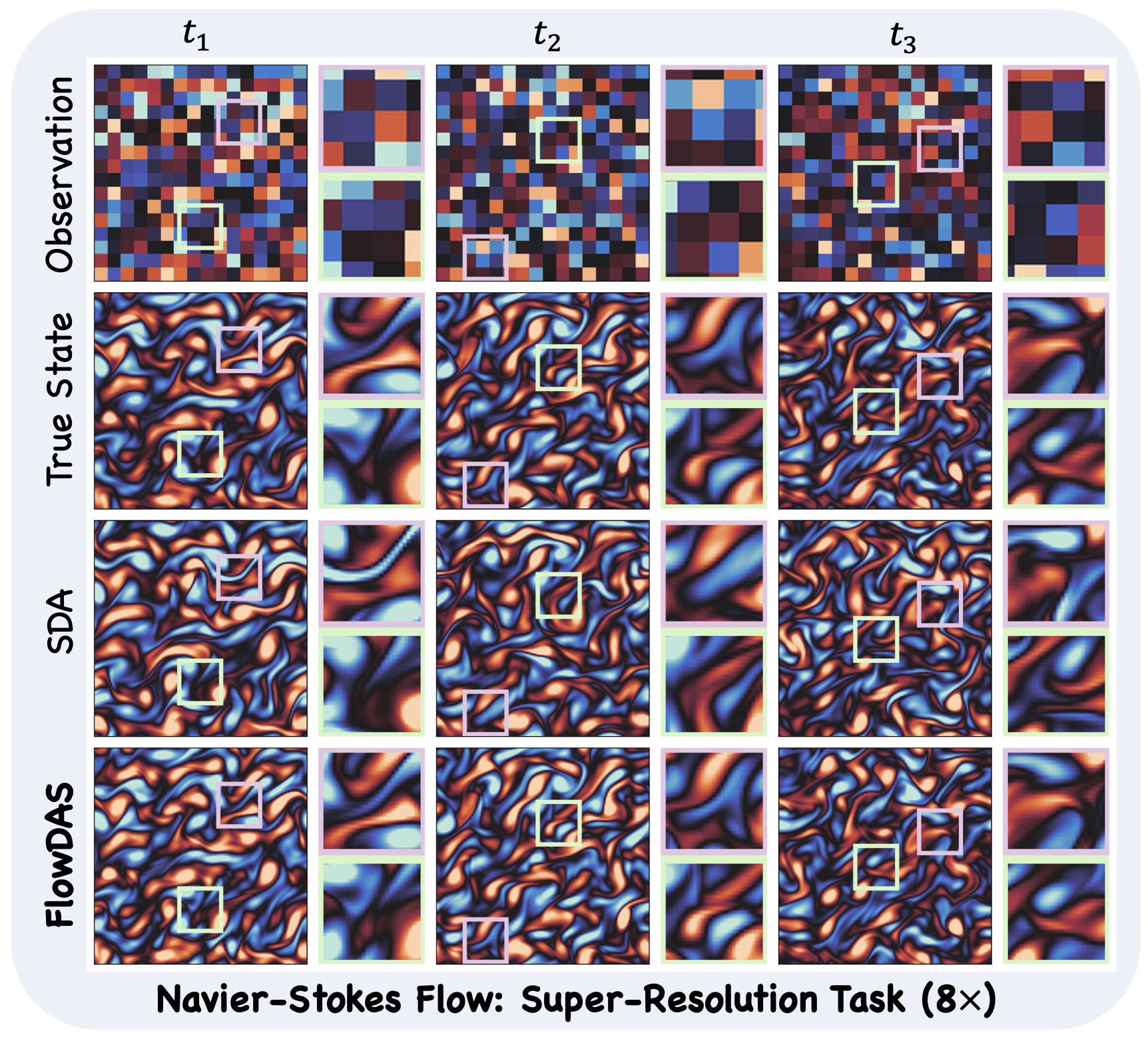}
    \caption{\textbf{Data assimilation of incompressible Navier-Stokes flow---Super-Resolution task ($8\times$)}. Experiments were conducted at observation resolution of $16^2$. The goal of super-resolution task is to reconstruct high-resolution vorticity data ($128^2$) from low-resolution observations.}
    \label{fig:NS_appendix_SR_8x}
\end{figure*}

\begin{figure*}[ht!]
    \centering
    \includegraphics[width=0.75\textwidth]{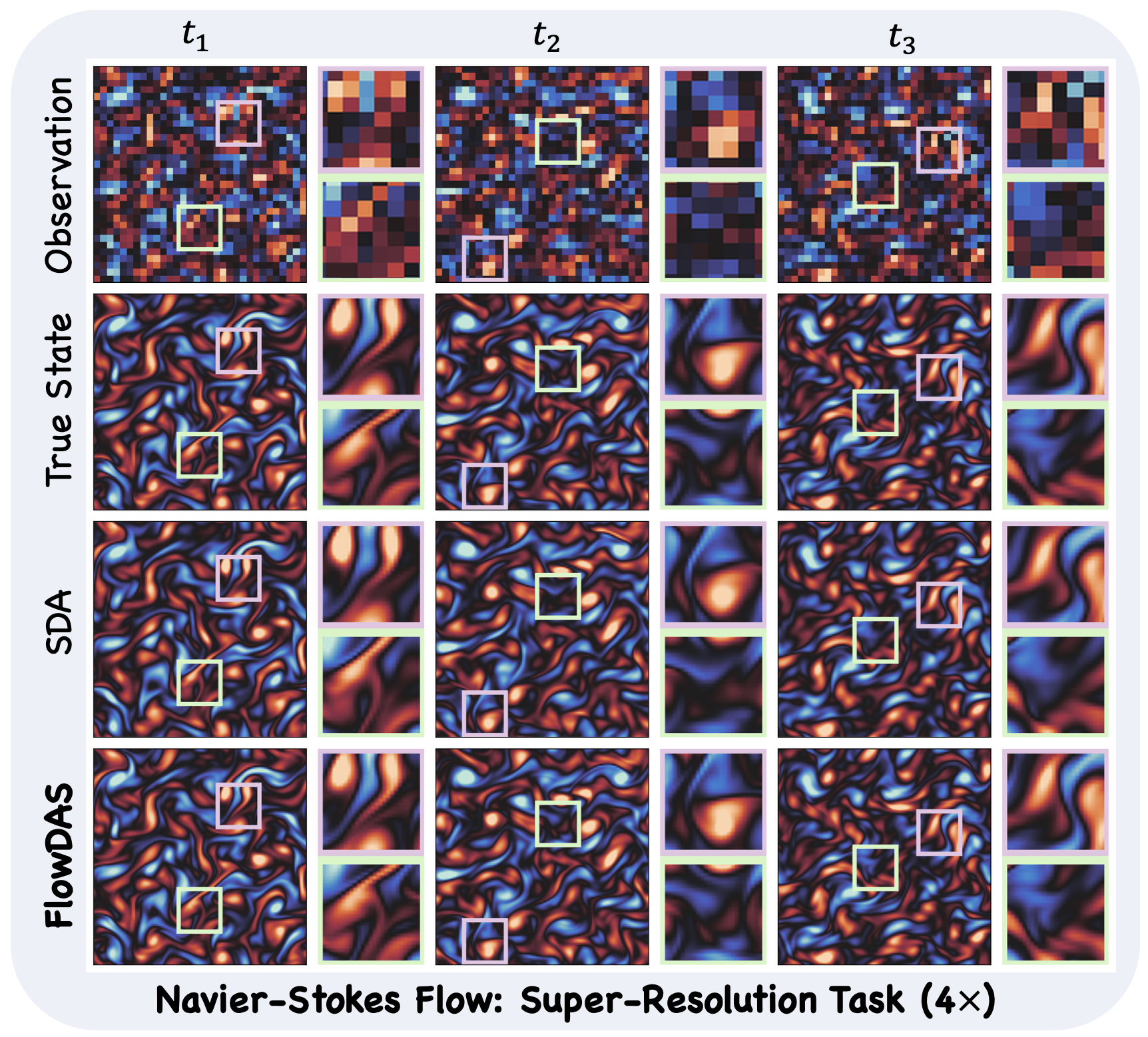}
    \caption{\textbf{Data assimilation of incompressible Navier-Stokes flow---Super-Resolution task ($4\times$)}. Experiments were conducted at observation resolution of $32^2$. The goal of super-resolution task is to reconstruct high-resolution vorticity data ($128^2$) from low-resolution observations.}
    \label{fig:NS_appendix_SR_4x}
\end{figure*}

\begin{figure*}[ht!]
    \centering
    \includegraphics[width=0.75\textwidth]{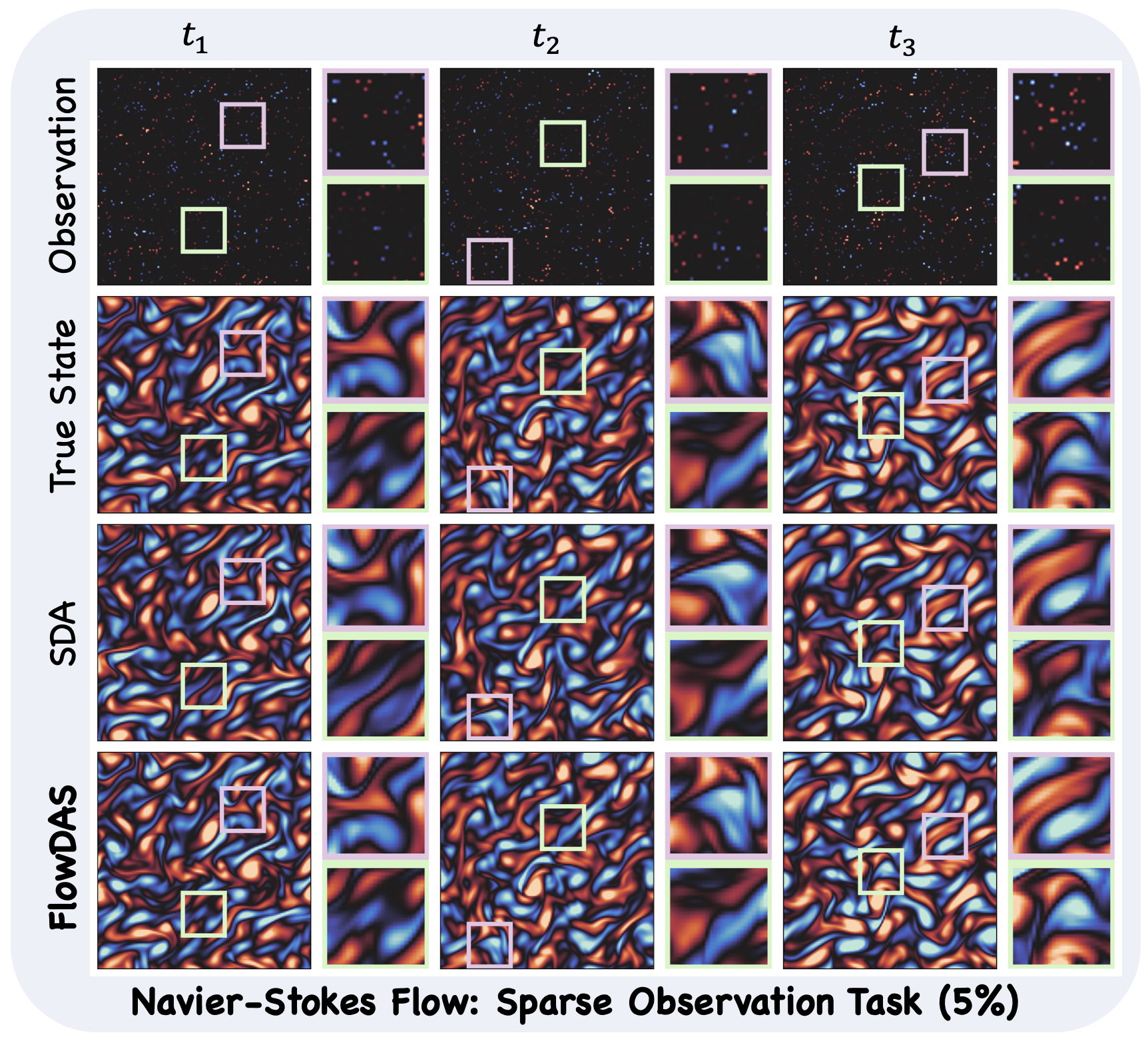}
    \caption{\textbf{Data assimilation of incompressible Navier-Stokes flow---Sparse-Observation task ($5\%$)}. Experiments were conducted at observation sparsity level at $5\%$. The goal of sparse observation task is to recover the complete vorticity field.}
    \label{fig:NS_appendix_SO_5}
\end{figure*}

\subsubsection{Additional Results}
\label{sec: supp on ns}
In this section, we present additional results to \Cref{sec:NS}, including the kinetic energy spectrum and visualization results for the Navier-Stokes flow super-resolution and sparse observation tasks.

\paragraph{\( 16^2 \rightarrow 128^2 \) Super-Resolution}
The kinetic energy spectrum is shown in \Cref{fig:spectral_4} (a), and additional visualization results are provided in \Cref{fig:NS_appendix_SR_8x}. FlowDAS effectively reconstructs high-resolution vorticity data from low-resolution observations, while SDA struggles to capture high-frequency physics.

\paragraph{\( 32^2 \rightarrow 128^2 \) Super-Resolution}
The kinetic energy spectrum is shown in \Cref{fig:spectral_4} (b), and additional visualization results are provided in \Cref{fig:NS_appendix_SR_4x}. Similar to the \( 16^2 \rightarrow 128^2 \) task, FlowDAS achieves superior alignment with the true state compared to SDA.

\paragraph{\( 5\% \) Sparse Observation}
The kinetic energy spectrum is presented in \Cref{fig:spectral_4} (d), with additional visualization results in \Cref{fig:NS_appendix_SO_5}. FlowDAS accurately recovers the vorticity field despite the highly sparse observations, significantly outperforming SDA, demonstrating strong recovery of physical information.

Across all tasks, FlowDAS consistently outperforms SDA, demonstrating superior alignment with the true kinetic energy spectrum and effective recovery of the vorticity field. FlowDAS not only excels in super-resolution tasks but also handles sparse observation challenges with robustness, maintaining physical coherence and accurately capturing high-frequency dynamics.


\subsection{Particle Image Velocimetry}
\label{sec:supp_on_PIV}
This section presents a realistic application of our method: Particle Image Velocimetry (PIV). PIV is a widely used optical technique for measuring velocity fields in fluids, with many scientific applications in aerodynamics \cite{taylor2010long, van2013piv, koschatzky2011high}, biological flow studies \cite{king2007piv, stamhuis2006basics, ergin2018review}, and medical research \cite{chen2014effects, tan2009analysis, ozcan2023integrated}.

\subsubsection{Task Description}
\begin{figure}
    \centering
    \includegraphics[width=0.75\linewidth]{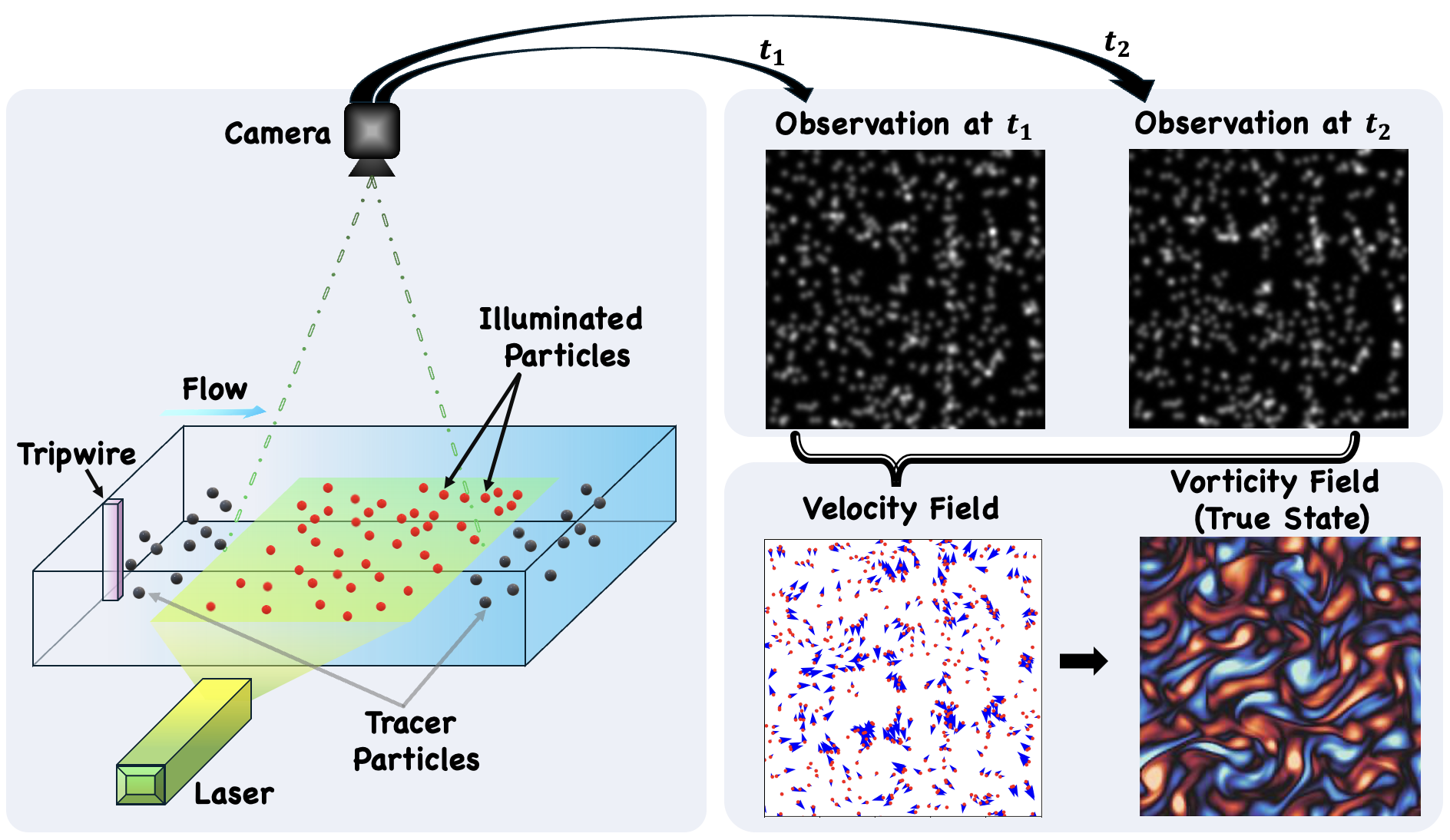}
    \caption{\textbf{Illustration of the real-world Particle Image Velocimetry (PIV) experiment}: The flow is seeded with tracer particles illuminated by a laser sheet, and their movements are captured by a camera to derive the sparse velocity field. The goal here is to recover dense vorticity field from the sparse velocity measurement.}
    \label{fig:piv_overview}
\end{figure}

In a standard PIV setup, as shown in \Cref{fig:piv_overview}, fluorescent tracer particles are seeded into a fluid flowing through a channel with transparent walls. A laser sheet illuminates the fluid, and particle movements are recorded by a high-speed camera with adjustable temporal resolution. By analyzing the displacement of these tracer particles, the velocity field within the fluid can be determined at sparse locations. Unlike the task in \Cref{sec:NS}, which involves recovering dense vorticity fields from sparse vorticity observations, PIV introduces a slightly different DA task: recovering dense vorticity fields ($\x = \bm \omega$) from sparse velocity measurements. This observation model is defined by
\begin{equation}
y = \cA (\blv(\bm \omega)) + \bm \eta,
\end{equation}
where the $\cA(\cdot)$ sparsely samples the velocity field \blv and the observation noise $\bm \eta$ has a standard deviation of $\gamma=0.25$. The relationship between the velocity \blv and the state (vorticity) $\bm \omega$ is given by $\bm \omega = \nabla \times \blv$. To derive the velocity $\bm{v}$ from the vorticity $\bm{\omega}$, we first solve the Poisson equation $\Delta \psi = -\bm{\omega}$ to obtain the stream function $\psi$, and then compute the velocity \blv as the gradient of $\psi$. This process is performed using the Fast Fourier Transform.

\subsubsection{Dataset}
In this experiment, we use the same fluid dynamics simulation data from the NS experiments described in \Cref{sec:NS}. However, we convert the vorticity data to velocity fields via Fourier transform to create synthetic PIV datasets \cite{Lagemann2021DeepRO}. The particle positions in our simulation are randomly initialized and then perturbed according to the simulated flow motion pattern. In these synthetic images, we assume a particle density of 0.03 particles per pixel, a particle diameter of 3 pixels, and a peak intensity of 255 for each particle in grayscale. The images are processed through a standard PIV pipeline to extract particle locations, match corresponding particles across frames, and compute sparse velocity observations. These sparse measurements are then used in DA to reconstruct the full vorticity fields.

\subsubsection{Baselines} 
We compare our method against the SDA solver. Instead of training new scores or stochastic interpolant networks, we directly adopt the networks trained on vorticity data from the incompressible NS flow experiment to evaluate FlowDAS and SDA on the PIV task.

\subsubsection{Experiment Setting}
The training data, neural network (including FlowDAS and SDA) is the same as those in the incompressible Navier-stokes flow simulation. \Cref{fig:piv_overview} shows the standard PIV set up. Hyperparameters of the inference procedure, specified for the PIV task, are presented in \Cref{table:hyperparameters}.

\begin{table}[t]
\begin{center}
\begin{sc}

\begin{tabular}{lcccccc}
\toprule
Parameter & Value & Unit \\
\midrule
Particle density  & 0.03 & particle per pixel \\
Particle diameter & 3 & Pixel \\
Peak intensity & 255 & Gray value \\
\bottomrule
\end{tabular}
\end{sc}
\end{center}
\caption{\textbf{PIV task: parameters for simulation.} This table summarizes parameters for the PIV experiments in detail.} 
\label{tab:PIV_paras}
\end{table}

\begin{figure*}[t]
\begin{center}
\centerline{\includegraphics[width=\textwidth]{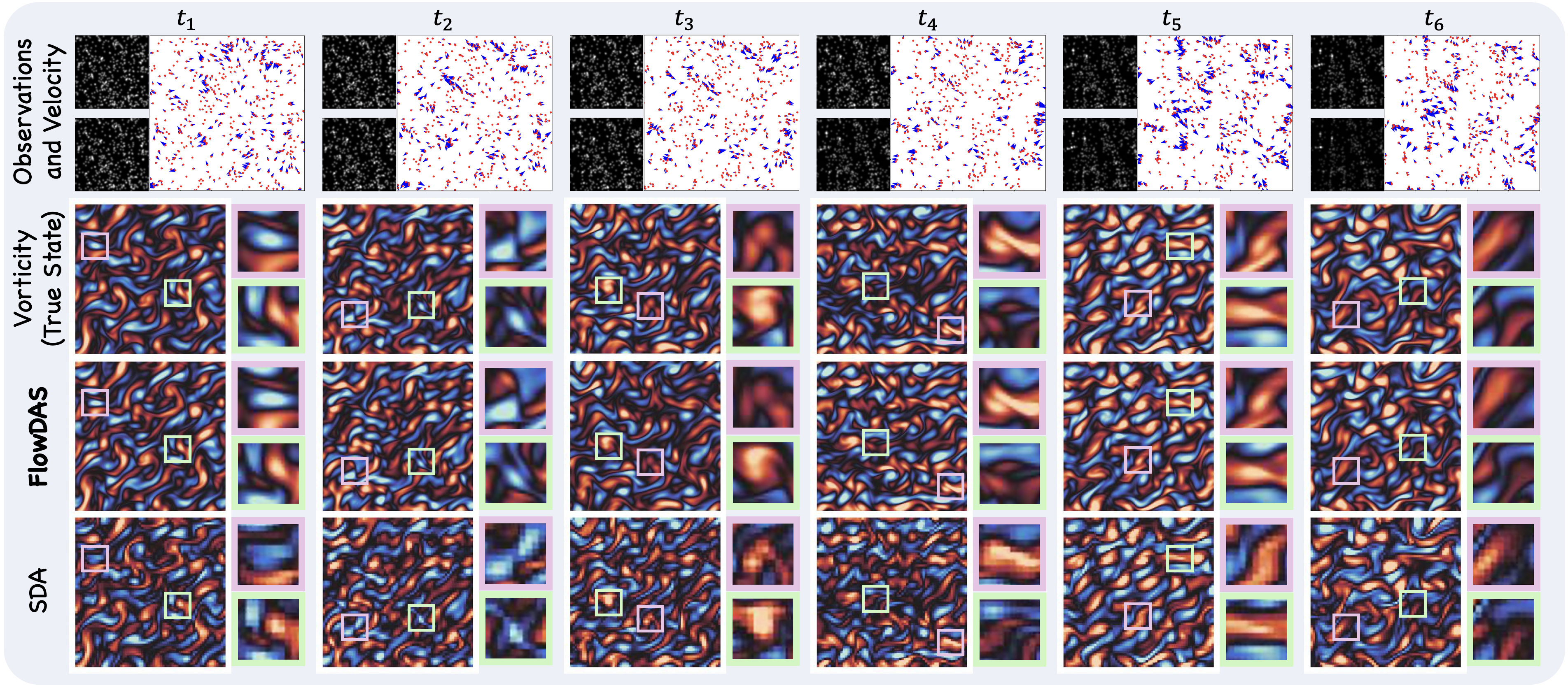}}
\caption{\textbf{Data assimilation of Particle Image Velocimetry.} 
The vorticity field is visualized in the same way as in \Cref{sec:NS}. FlowDAS outperforms SDA in terms of reconstruction fidelity and RMSE, recovering more detailed features even when direct observations are not available. These improvements highlight the potential of FlowDAS for real-world applications.}
\label{fig:PIV_results}
\end{center}
\end{figure*}

\subsubsection{Results}
FlowDAS accurately reconstructs the underlying fluid dynamics from observed particle images, producing vorticity estimates with high precision. As shown in \Cref{fig:PIV_results}, it outperforms the baseline SDA in terms of reconstruction fidelity. Quantitatively, FlowDAS achieves a lower average RMSE of 0.118, compared to 0.154 for SDA. This performance gap highlights the robustness of FlowDAS and its potential utility in fluid dynamics applications.

\subsection{Weather Forecasting}
\label{sec:supp_on_SEVIR}

\subsubsection{Baselines} 
We compared FlowDAS to FNO-DA, Transolver-DA, and SDA. 

\paragraph{FNO-DA}
The model is configured with $7$ stacked Fourier layers. Each layer processes data in a $128$-dimensional feature space and truncates the Fourier series to include 64 modes in each spatial direction. A padding of $6$ is applied before the Fourier operations and removed afterward. The model is provided with the previous $6$ states to predict the state at the next time step. We train the model for $1000$ epochs, using a batch size of $200$ and the AdamW optimizer with an initial learning rate of $0.0001$.

\paragraph{Transolver-DA} 
The model has $8$ layers of transolver blocks whose hidden channels are $128$ and have $32$ slices, with $8$ attention heads. The model is provided with the previous $6$ states to predict the state at the next time step. We train the model for $1000$ epochs using a batch size of $200$, the AdamW optimizer, and a OneCycleLR learning rate scheduler with an initial learning rate of  $5e^{-4}$.

\paragraph{SDA}
The score neural network is configured with a temporal window of $7$, of which the first $6$ inputs are conditions, and the forecast horizon is $1$; an embedding layer dimensionality of $32$; and a base number of feature channels $256$. The network depth is $5$, and the activation function used is SiLU \cite{Loshchilov2017DecoupledWD}. We train the model using a batch size of $200$ and the AdamW optimizer, with a linear learning rate scheduler (initial learning rate $0.001$) and a weight decay of $0.001$.

During inference, the $\alpha_1$ and $\alpha_2$ used is varied from task to task for both the FNO-DA and the Transolver-DA. We show the detailed parameter settings in \Cref{table:hyperparameters 4 transolver,table:hyperparameters 4 FNO}. We use an AdamW optimizer for Kalman update with a learning rate of $1e^{-4}$ and a maximum iteration time of $2000$ to ensure the loss finally becomes stable. Without other indications, we adopt the same optimizer settings for both FNO-DA and Transolver-DA.

\subsubsection{Training Details} We experiment with three backbone architectures---U-Net, FNO, and Transolver---to learn the drift term and compare their performance and provide numerical results in \Cref{tab:ablate_drift}. The architectural details are as follows:
\begin{itemize}[leftmargin=*]
    \item UNet: as shown in \Cref{subsub sec:details about the ns neural network}, the only change is the number of states conditioned on. Here we condition on previous $6$ states.
    \item FNO: this model is configured with $7$ stacked Fourier layers. Each layer processes data in a $128$-dimensional feature space and truncates the Fourier series to include $64$ modes in each spatial direction. A padding of $6$ is applied before the Fourier operations and removed afterward. The model is provided with the previous $6$ states, the $\Xs$ and the time $s$ as additional channel concatenated to the previous states and the $\Xs$ to predict the drift term. 
    \item Transolver: this model has $5$ layers of transolver blocks whose hidden channels are $128$ and have $32$ slices, with $8$ attention heads. The sinusoidal timestep embeddings are applied. The inputs are the previous $6$ states, the $\Xs$ and the time $s$ and the output is the drift term.
\end{itemize}

\begin{table}[t]
\begin{center}
\begin{sc}

\begin{tabular}{lccc}
\toprule
\textbf{Backbone of FlowDAS} & \textbf{RMSE} $\downarrow$ & \textbf{CSI($\tau_{20}$) (0.3)} $\uparrow$ & \textbf{CSI($\tau_{40}$) (0.5)} $\uparrow$\\
\midrule
FNO         & \textbf{0.053$\pm$0.001} & \textbf{0.718$\pm$0.015} & \textbf{0.568$\pm$0.024} \\
Transolver  & 0.056$\pm$0.002 & 0.702$\pm$0.018 & 0.540$\pm$0.028 \\
UNet        & 0.056$\pm$0.002 & 0.703$\pm$0.017 & 0.540$\pm$0.023 \\
\bottomrule
\end{tabular}

\end{sc}
\end{center}
\caption{Comparison of backbone network architectures of the drift model on the weather forecasting task. All metrics are averaged over multiple runs. Subtle improvements are observed when using a FNO network as the backbone of the drift model.}
\label{tab:ablate_drift}
\end{table}

\subsubsection{Results}
\begin{figure}
    \centering
    \includegraphics[width=0.9\linewidth]{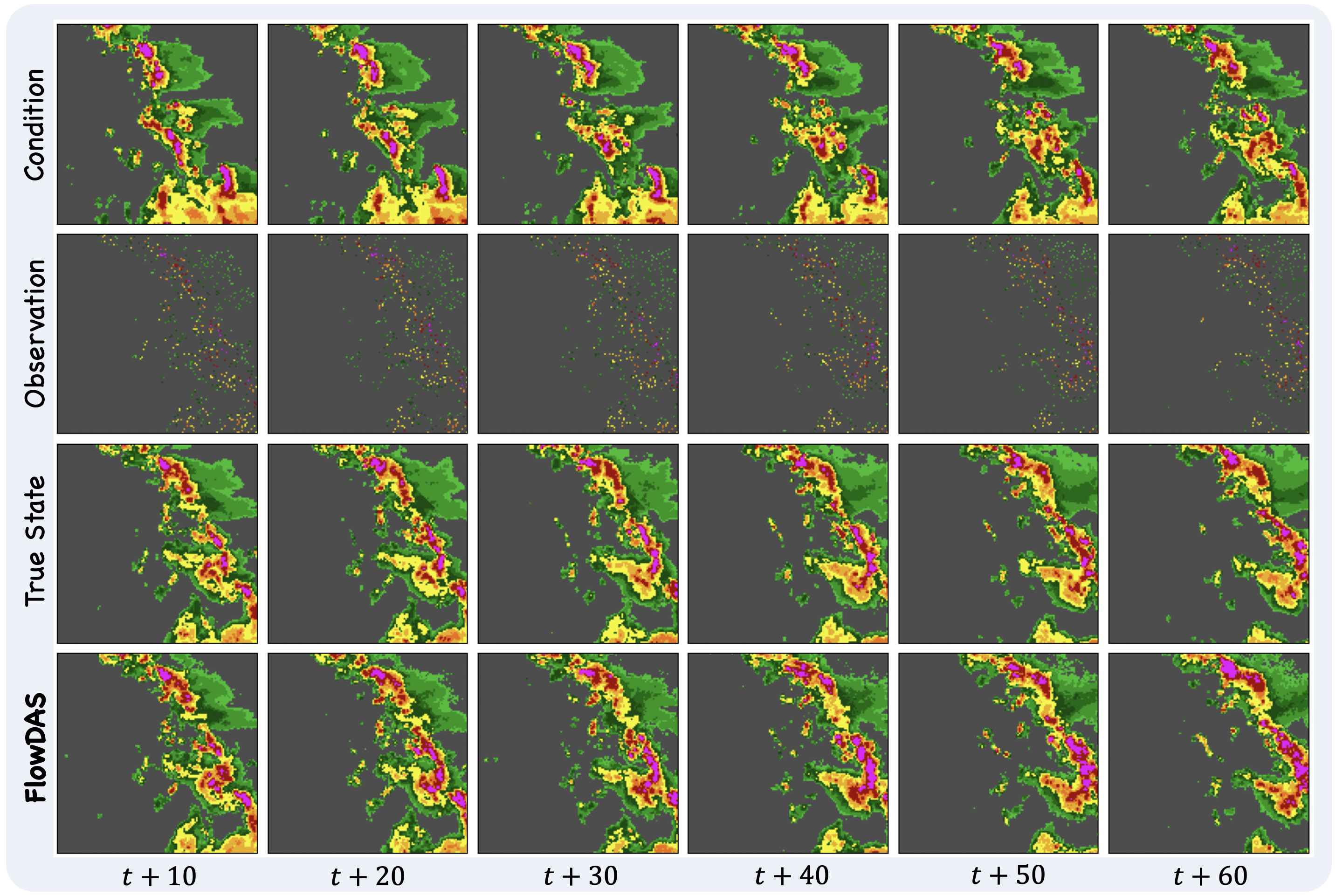}
    \caption{\textbf{Additional result 1: Data assimilation of weather forecasting (Sparse-Observation task)}. The underlying PDE of the dynamical system is unknown. This experiment was conducted at an observation sparsity level of \(10\%\), showing the capacity of long-trajectory forecasting.}
    \label{fig:sevir1}
\end{figure}

\begin{figure}
    \centering
    \includegraphics[width=0.9\linewidth]{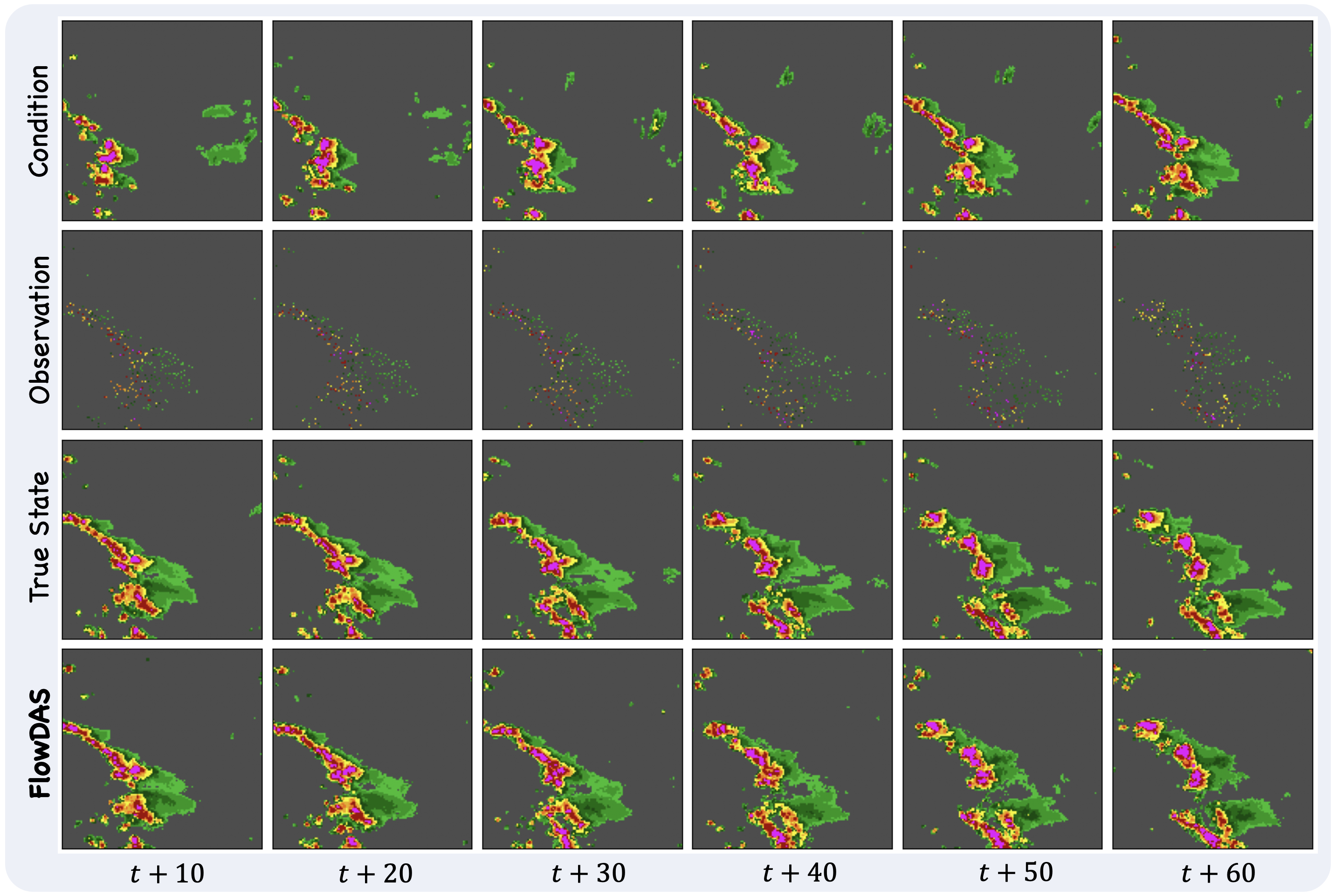}
    \caption{\textbf{Additional result 2: Data assimilation of weather forecasting (Sparse-Observation task)}. The underlying PDE of the dynamical system is unknown. This experiment was conducted at an observation sparsity level of \(10\%\), showing the capacity of long-trajectory forecasting.}
    \label{fig:sevir2}
\end{figure}

\begin{figure}
    \centering
    \includegraphics[width=0.9\linewidth]{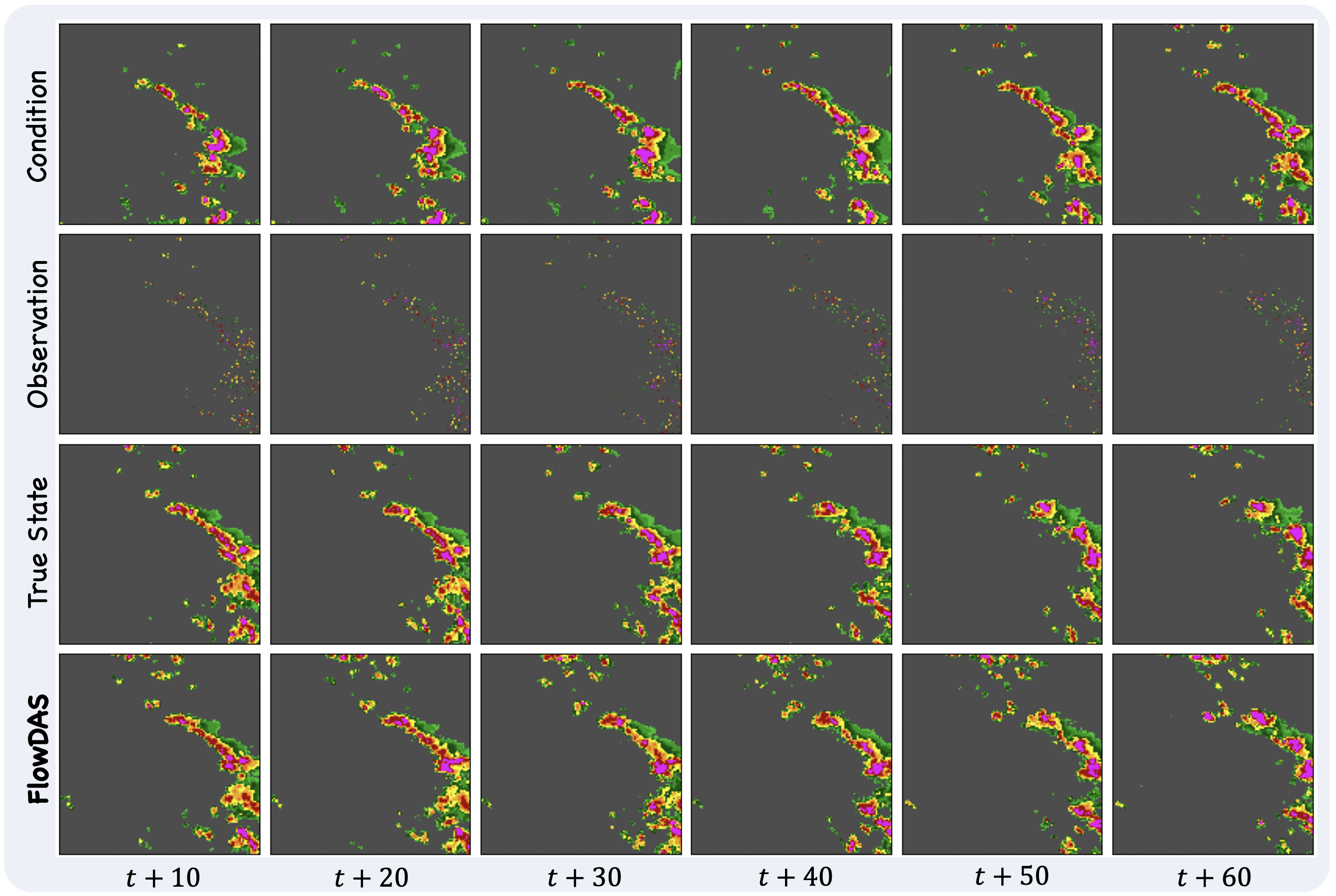}
    \caption{\textbf{Additional result 3: Data assimilation of weather forecasting (Sparse-Observation task)}. The underlying PDE of the dynamical system is unknown. This experiment was conducted at an observation sparsity level of \(10\%\), showing the capacity of long-trajectory forecasting.}
    \label{fig:sevir3}
\end{figure}

We present additional visualizations to demonstrate FlowDAS's ability of capturing unknown system dynamics on this weather forecasting task in \Cref{fig:sevir1,fig:sevir2,fig:sevir3}. FlowDAS successfully generates accurate estimates of the weather in the future frames in an autoregressive manner. \Cref{tab:sevir} provides numerical results of FlowDAS as well as other baseline methods. FlowDAS consistently outperformed DA baselines both in terms of RMSE and CSI metrics. 

\addtocontents{toc}{\protect\setcounter{tocdepth}{2}}  

\end{document}